\documentclass[11pt,a4paper]{article}
\pdfoutput=1
\usepackage{jheppub}
\usepackage{rotating}
\usepackage{array}
\usepackage{relsize} 
\usepackage{amsmath}
\usepackage{slashed}
\usepackage{booktabs}
\usepackage[pdftex,table]{xcolor}
\usepackage{mathtools}
\usepackage{empheq}
\usepackage{float}
\usepackage{blindtext}
\usepackage{wasysym} 

\usepackage{etoolbox}
\makeatletter
\patchcmd{\CatchFBT@Fin@l}{\endlinechar\m@ne}{}
  {}{\typeout{Unsuccessful patch!}}
\makeatother

\newcommand{\di}{\text{d}}

\newcommand{\eqsp}{\;}

\DeclareMathOperator{\acoth}{acoth}

\preprint{ULB-TH/21-14}

\title{Dark matter produced from neutrinos}

\author[a]{Marco Hufnagel}
\author[a,b]{and Xun-Jie Xu}

\affiliation[a]{Service de Physique Théorique, Université Libre de Bruxelles, Boulevard du Triomphe, CP225, B-1050 Brussels, Belgium}
\affiliation[b]{Institute of High Energy Physics, Chinese Academy of Sciences, Beijing 100049, China}

\abstract{
	In the presence of interactions between neutrinos and dark matter (DM), DM can potentially be produced via freeze-in from  the  neutrino  sector. We investigate the  implications of such a scenario for the evolution of both DM and neutrinos in the early Universe, and show that the future cosmic neutrino detection experiment \textsc{PTOLEMY} might be sensitive to neutrino signals that originate from DM annihilation in this model.
}

\begin{document}
\maketitle

\section{Introduction}
Since both neutrinos and dark matter (DM) are two of the most compelling pieces of evidence for new physics beyond the Standard Model (BSM), it is conceivable that both particles are indeed connected and thus feature non-trivial interactions among each other. Such a scenario could then potentially induce noticeable changes to the evolution of the Universe~\cite{Serra:2009uu,Mangano:2006mp,Wilkinson:2014ksa,Bertoni:2014mva,Berlin:2017ftj,DiValentino:2017oaw,Olivares-DelCampo:2017feq,Berlin:2018ztp,Stadler:2019dii,Sabti:2019mhn,Depta:2019lbe,Becker:2020hzj,Mosbech:2020ahp,Paul:2021ewd,Green:2021gdc}, as DM-neutrino interactions are capable of changing the anisotropies in the cosmic microwave background (CMB)~\cite{Serra:2009uu,Wilkinson:2014ksa,Becker:2020hzj,Mosbech:2020ahp}, modifying big bang nucleosynthesis (BBN)~\cite{Sabti:2019mhn,Depta:2019lbe}, and/or affecting structure formation at small scales~\cite{Mangano:2006mp,Bertoni:2014mva}.
Additionally, one might ask the question if interactions between DM and neutrinos could also be responsible for setting the correct relic abundance, e.g.~via freeze-out~\cite{Kolb} or freeze-in~\cite{Hall:2009bx} of DM from the neutrino bath. In the former case, the final relic abundance mainly depends on the thermally averaged cross-section of the annihilation process\footnote{For specific formulae see e.g.~\cite{Plehn:2017fdg,Zyla:2020zbs}.} and the freeze-out temperature is typically above or close to the neutrino decoupling temperature $T_{\nu\text{-dec}}\approx~1.4\,\mathrm{MeV}$~\cite{Bennett:2019ewm}, in order to comply with stringent bounds from BBN~\cite{Depta:2019lbe}. Consequently, this production mechanism can be covered with the usual freeze-out formalism. In the latter case, however, the physical process can freely happen both before or after neutrino decoupling, and the results are usually more process dependent.

In this work, we consider this exact scenario, i.e. the possibility that DM is produced via freeze-in from the neutrino sector, and we investigate the implications of this setup for the evolution of both DM and neutrinos in the early Universe. To this end, we solve the Boltzmann equation for the DM abundance not only at the level of the number density, but also at the level of the spectrum, with the latter one being crucial for the study of Lyman-$\alpha$ constraints.

Moreover, a particularly noteworthy feature of our scenario is the fact that DM annihilation, happening in the galactic center at the present time, can lead to monochromatic neutrino lines, which -- if detected -- would constitute a ``smoking-gun'' signal for DM detection. Previous studies~\cite{Dudas:2014bca,ElAisati:2015qec,ElAisati:2015ugc,Garcia-Cely:2017oco,Coy:2020wxp,Coy:2021sse} have already shown that neutrino lines from DM decays can potentially be observed in neutrino detectors such as \textsc{Super-K}~\cite{Super-Kamiokande:2002exp}, \textsc{KAMLAND}~\cite{KamLAND:2011bnd}, or \textsc{BOREXINO}~\cite{Borexino:2010zht}, whereas DM annihilation is generally considered to be too weak to allow for realistic detection~\cite{ElAisati:2017ppn}. However, in this work, we show that within the sensitivity of the proposed \textsc{PTOLEMY}~\cite{PTOLEMY:2018jst,PTOLEMY:2019hkd} experiment, which aims at the detection of the cosmic neutrino background (C$\nu$B), DM annihilation could indeed be detected.

Our work is structured as follows: In sec.~\ref{sec:Framework}, we first introduce two different models with DM-neutrino interactions and present the relevant matrix elements for both scenarios. In sec.~\ref{sec:Evolution}, we then study the evolution of the new particles in these models by solving the respective Boltzmann equations, both for the number density as well as for the spectrum. In sec.~\ref{sec:Constraints}, we then present the different constraints that are imposed on our model, and in sec.~\ref{sec:nu_signal}, we study the neutrino signal from DM annihilation in the galactic center and its detection prospects with the \textsc{PTOLEMY} experiment. Finally, we conclude in sec.~\ref{sec:Conclusion}, while some details of our calculations are delegated to the appendix.

\section{Framework}
\label{sec:Framework}

In this work, we extend the Standard Model (SM) by a dark sector that includes a DM particle $\chi$ as well as a mediator $\phi$. Within this setup, we then study the production of DM from SM neutrinos via reactions of the form $\nu \bar{\nu} \rightarrow \chi \bar{\chi}$, which can happen via either $t$- or $s$-channel mediator exchange.

\subsubsection*{Lagrangians}

To implement the $\boldsymbol{t}$\textbf{-channel} scenario, we assume that the mediator is a complex scalar boson $\phi$, which couples to the SM neutrinos and the DM particle via the interaction Lagrangian
\begin{equation}
{\cal L}_t\supset y_{\chi\nu}{\sum}_i\phi\bar{\chi}P_{L}\nu_i+{\rm h.c.}\eqsp,
\label{eq:lag_t}
\end{equation}
where the left-handed projector $P_{L}=(1-\gamma^{5})/2$ ensures that only left-handed neutrinos are participating in the interactions, and the sum $\sum_i$ goes over all participating neutrino flavors. For now, we do not make any assumption regarding the number of active neutrinos that take part in the interaction, but we will later limit our considerations to only one participating neutrino at a time. Models like this are mainly motivated by previous studies on DM-neutrino interactions from a neutrino portal~\cite{Orlofsky:2021mmy,Berryman:2017twh,Becker:2018rve,Batell:2017rol,Falkowski:2009yz}, which often feature a scalar that is charged under a hidden symmetry in order to guarantee DM stability.

\begin{figure}[t]
	\centering
	\includegraphics[width=0.95\textwidth]{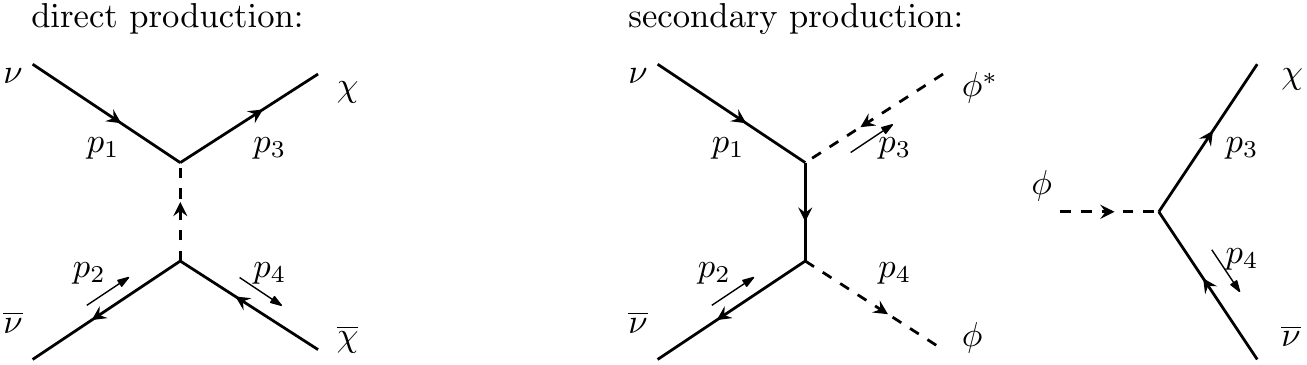}
	\caption{Feynman diagrams for the production of $\chi$ (left) and $\phi$ (center) from neutrinos, as well as the diagram for $\phi$ decay (right), which can potentially lead to secondary DM production.}
	\label{fig:diagrams_1}
\end{figure}

While the Lagrangian in eq.~\eqref{eq:lag_t} does enable the intended reaction $\nu \bar{\nu} \leftrightarrow \chi \bar{\chi}$, it also induces the additional production process $\nu \bar{\nu} \leftrightarrow \phi\phi^*$, which might lead to secondary DM production via the decay of the mediator, as well as the two (inverse) decay channels $\phi \leftrightarrow \chi \bar{\nu}$ if $m_\phi > m_\chi$ and $\chi \leftrightarrow \phi \bar{\nu}$ if $m_\phi < m_\chi$. In order for $\chi$ to be stable, we thus enforce the additional condition $m_\phi > m_\chi$, which renders the latter decay channel irrelevant. Consequently, there remain three relevant processes in this scenarios, which correspond to the Feynman diagrams that are depicted in fig.~\ref{fig:diagrams_1}.

\begin{figure}[t]
	\centering
	\includegraphics[width=0.95\textwidth]{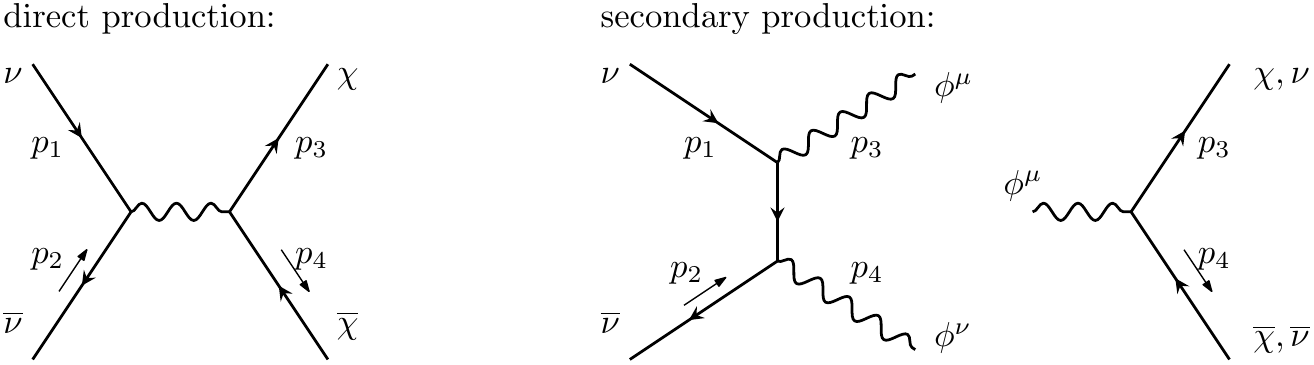}
	\caption{Similar to fig.~\ref{fig:diagrams_2}, but for the $s$-channel case.}
	\label{fig:diagrams_2}
\end{figure}

For the $\boldsymbol{s}$\textbf{-channel} scenario, we instead assume that the mediator is a real vector boson $\phi_\mu$, which couples separately to DM and neutrinos via terms of the form
\begin{equation}
{\cal L}_s \supset y_{\chi}\phi_{\mu}\bar{\chi}\gamma^{\mu}P_{L}\chi + y_{\nu}{\sum}_i\phi_{\mu}\bar{\nu}_i\gamma^{\mu}P_{L}\nu_i\eqsp.\label{eq:lag_s}
\end{equation}
Such interactions frequently occur in models with kinetic mixing between the SM gauge bosons and the one of a new $U(1)_X$ gauge group~\cite{Holdom:1985ag}.
Alternatively, it would also be possible to consider a scalar mediator with lepton-number violating interactions, analogous to the Majoron~\cite{Chikashige:1980ui}. However, we do not discuss such a scenario here, and only note that both scenarios would lead to rather comparable results. Coming back to our scenario, the Lagrangian in eq.~\eqref{eq:lag_s} again enables the desired reaction $\nu \bar{\nu} \leftrightarrow \chi \bar{\chi}$, but also the additional processes $\nu \bar{\nu} \leftrightarrow \phi \phi$, $\phi \leftrightarrow \nu \bar{\nu}$, and, if $m_\phi > 2m_\chi$, $\phi \leftrightarrow \chi \bar{\chi}$. The latter reactions can again lead to secondary DM production like in the $t$-channel case. For our purposes, it is however possible to neglect the secondary production processes by concentrating on $m_\phi < 2m_\chi$ (cf.~section~\ref{sec:nu_signal}), in which case\footnote{Note that in the $s$-channel scenario $m_\phi< m_\chi$ is possible, since $\phi$ is always allowed to decay into neutrinos, meaning that it can never be a viable DM candidate.} $\phi$ can only decay into neutrinos and therefore does not alter the final DM abundance.
In addition, it is worth noting that we only consider scenarios in which the mediator abundance is small compared to the one of neutrinos. Consequently, its production also does not significantly distort the neutrino spectrum, meaning that its influence is ultimately negligible. Hence, we end up with only a single relevant process, namely the leftmost Feynman diagram in fig.~\ref{fig:diagrams_2}.


\subsubsection*{Matrix elements}

Given these two Lagrangians, the squared matrix elements for the relevant production processes can be calculated with
\textsc{FeynCalc}~\cite{Shtabovenko:2020gxv}
or \textsc{Package-X}~\cite{Patel:2015tea}.
In the case of $\boldsymbol{t}$\textbf{-channel} annhilations according to eq.~\eqref{eq:lag_t}, the transition amplitudes for the production of $\chi$'s and $\phi$'s from neutrinos are given by
\begin{align}
|{\cal M}^{(t)}_{\nu\bar{\nu}\rightarrow\chi\bar{\chi}}|^{2}  =N_\nu y_{\chi\nu}^{4}\left(\frac{t-m_{\chi}^{2}}{t-m_{\phi}^{2}}\right)^{2}\eqsp \quad\text{and}\quad |{\cal M}^{(t)}_{\nu\bar{\nu}\rightarrow\phi\phi^*}|^{2} = N_\nu y_{\chi\nu}^{4}\frac{tu - m_\phi^4}{(t-m_\chi^2)^2}\eqsp,
\label{eq:matrix_t}
\end{align}
respectively. Here, $t=(p_{3}-p_{1})^{2}$, $s=(p_{1}+p_{2})^{2}$, and $u=(p_{1}-p_{4})^{2}$ are the Mandelstam variables, and $N_\nu \in \{1, 2, 3\}$ is the number of participating neutrinos. For the secondary production of $\chi$ via the decay of $\phi$, we further need to know the rate of the process $\phi\rightarrow \chi \bar{\nu}$, which is given by
\begin{align}
\Gamma^{(t)}_{\phi\rightarrow\chi\bar{\nu}} = \frac{N_\nu y_{\chi\nu}^{2}\left(m_{\phi}^{2}-m_{\chi}^{2}\right){}^{2}}{16\pi m_{\phi}^{3}}\eqsp.
\end{align}

Given the Lagrangian from eq.~\eqref{eq:lag_s} with an $\boldsymbol{s}$-\textbf{channel} mediator, the matrix element for neutrinos annihilating into DM is given by
\begin{align}
|{\cal M}^{(s)}_{\nu\bar{\nu} \rightarrow \chi\bar{\chi}}|^{2} = \frac{4N_\nu y_{\chi}^{2}y_{\nu}^{2}\left(u-m_{\chi}^{2}\right)^{2}}{\left(s-m_{\phi}^{2}\right)^{2}+\left(m_{\phi}\Gamma_{\phi}^{(s)}\right)^2}
\label{eq:matrix_s}
\end{align}
with the total decay width $\Gamma_\phi^{(s)}$ of $\phi$. For $m_\phi < 2m_\chi$, the only contribution to the latter quantity comes from the decay channel $\phi\rightarrow\nu\bar{\nu}$,
and is given by
\begin{equation}
\Gamma^{(s)}_{\phi} = \Gamma^{(s)}_{\phi\rightarrow \nu\bar{\nu}}=\frac{N_\nu y_{\nu}^{2}}{24\pi}m_{\phi}\eqsp.
\end{equation}

Given the above matrix elements, both scenarios can be described by the mass $m_\chi$, the ratio $m_\phi/m_\chi$, the coupling
\begin{align}
y \equiv \begin{cases}
y_{\chi \nu} &\qquad t\text{-channel} \\
\sqrt{y_\chi y_\nu} &\qquad s\text{-channel}
\end{cases}\eqsp,
\label{eq:m-55}
\end{align}
and, additionally in case of an $s$-channel mediator, the lifetime $\tau_\phi^{(s)}=1/\Gamma_{\phi}^{(s)}$~. However, we checked that the lifetime only has an effect on our results if $\tau_\phi \lesssim 10^{-21}\,\mathrm{s}\times(1\,\mathrm{MeV}/m_\phi)$, which corresponds to large couplings $y_\nu\gtrsim 7.04/\sqrt{N_\nu}$ above the pertubativity limit. We are thus not interested in couplings of this magnitude and therefore simply set $\tau_\phi \rightarrow \infty$ for the remainder of this work, corresponding to $y_\nu \rightarrow 0$. In a next step, we now strive to solve the appropriate Boltzmann equation for the freeze-in process in order to calculate the resulting abundance of $\chi$ for each combination of parameters $(m_\chi, m_\phi/m_\chi, y)$. This procedure then allows us to fix $y$ in such a way to obtain the observed DM relic abundance, which only leaves us with two free parameters in both scenarios.

\section{Thermal evolution of the dark sector}
\label{sec:Evolution}

\subsection{The number densities of DM and the mediator}
\label{sec_evo_number}
In this work, we focus on the case where the produced DM and mediator abundances are small compared to the one of thermal neutrinos (which is true roughly until matter-radiation equality), as this allows us to safely neglect the inverse reactions $\chi\bar{\chi}\rightarrow \nu \bar{\nu}$ and $\phi\phi^*\rightarrow \nu \bar{\nu}$, while also ensuring that the neutrinos approximately retain their Fermi-Dirac distribution throughout the full production process. Then, when taking into account \emph{only} the relevant annihilation reactions (we will discuss the handling of the decay below), the number density of each dark-sector particle $X \in \{\chi, \phi\}$ with anti-particle $\bar{X} \in \{\bar{\chi}, \phi^*\}$\footnote{Here, $n_X$ is the number density of only $X$, but not of its anti-particle $\bar{X}$, and $\phi = \phi^*$ for the $s$-channel scenario.} evolves according to the integrated Boltzmann equation
\begin{align}
\dot{n}_X + 3Hn_X = \mathfrak{C}_{\nu \bar{\nu} \rightarrow X \bar{X}} \eqsp.
\label{eq:beq_n}
\end{align}
Here, the sum goes over all participating neutrinos, and the collision operator $\mathfrak{C}_{\nu \bar{\nu} \rightarrow X \bar{X}}$ describing the annihilation process is given by
\begin{align}
\mathfrak{C}_{\nu \bar{\nu} \rightarrow X \bar{X}} \simeq \frac{1}{32\pi^4} \int_0^\infty \text{d}E_1 \int_{m_X^2/E_1}^\infty \hspace{-0.3cm}\text{d} E_2 \; \frac{1}{e^{E_1/T_\nu}+1} \frac{1}{e^{E_2/T_\nu}+1} \int_{4m_\chi^2}^{4E_1 E_2} \hspace{-0.2cm}\text{d} s \; s\cdot\sigma_{\nu \bar{\nu} \rightarrow X \bar{X}}(s)\eqsp.
\end{align}
Here, the only approximation that we made was to neglect the spin-statistical factors $1-f_X$ for the dark-sector particles in the final state, which is justified since we assume the produced DM/mediator abundance to be small compared to the one of neutrinos, implying $f_X \ll \bar{f}_\nu\sim\bar{f}_X\sim1$ with the thermal distributions $\bar{f}_\nu$ and $\bar{f}_X$ of $\nu$ and $X~\in~\{\chi, \phi\}$, respectively. For both scenarios, the relevant cross-sections can be calculated from the matrix elements in eqs.~\eqref{eq:matrix_t} and \eqref{eq:matrix_s}, and we provide the relevant expressions in appendix~\ref{sec:M2}.

Properly handling the potential secondary production of DM particles via the decay of $\phi$ at the level of the number density is a rather futile endeavor, however, as the relevant collision operator explicitly involves the phase-space distribution $f_\phi$ of $\phi$, which cannot be deduced from $n_\phi$. Nevertheless, for the masses considered in this work, secondary production and hence the decay of the mediator is only relevant in the $t$-channel scenario,\footnote{Remember that, in the $s$-channel scenarios, we limit our considerations to the case $m_\psi < 2m_\chi$, which implies that $\phi$ can only decay into neutrinos.} in which case we find that the corresponding lifetime is short compared to the production time, e.g.~$\tau_\phi^{(t)} = 1/\Gamma^{(t)}_{\phi\rightarrow\chi\bar{\nu}} \sim 10^{-4}\,\mathrm{s}\times(1\,\mathrm{MeV}/m_\chi)$ for $y_{\chi\nu} = 10^{-6}$ and a rather tuned mass splitting $m_\phi/m_\chi = 1.01$. Hence, for the calculation of the final DM abundance, we can simply \textit{(1)} calculate the production of both $\chi$ and $\phi$ separately without the decay and \textit{(2)} set $n_\chi \rightarrow n_\chi + n_\phi$ after the production process of both particles has concluded. Using this approximation it is then sufficient to solve eq.~\eqref{eq:beq_n}, which yields
\begin{align}
n_X(t)R(t)^3 = \int_{t_i}^{t} \di t'\; \mathfrak{C}_{\nu \bar{\nu} \rightarrow X\bar{X}}(t') R(t')^3
\end{align}
with the scale factor $R$ and the initial conditions $n_X(t_i) = 0$ at $T_i \equiv T(t_i) \gg m_\chi, m_\phi$, for which we explicitly set $T_i = 10^3 \max\{m_\chi, m_\phi\}$.

In our setup, we can safely assume that freeze-in for both particles has concluded at $T_\text{eof} = 10^{-2}\min\{m_\chi, m_\phi\}$ corresponding to some time $t_\text{eof}$ with $T(t_\text{eof}) = T_\text{eof}$, after which the number densities of both particles remains only subject to redshift. Hence, $n_X R^3 = \text{const}$ for $T<T_\text{eof}$ and by setting
\begin{align}
n_{\chi*} \equiv \begin{cases}
n_\chi + n_\phi&\qquad t\text{-channel} \\
n_\chi &\qquad s\text{-channel}
\end{cases}\eqsp,
\label{eq:def_ns}
\end{align}
to incorporate the decay of $\phi$ into the abundance of $\chi$, the final relic abundances becomes
\begin{align}
\Omega_\chi h^2 = \frac{2m_\chi n_{\chi*}(t_\text{eof})R(t_\text{eof})^3}{\rho_{\text{crit}, 0} R_0^3/h^2}\eqsp,
\end{align}
where $\rho_{\text{crit}}$ is the critical energy density, $h$ is the Hubble rate in units of $100\,\mathrm{Mpc}/(\mathrm{km}\cdot\mathrm{s})$, and all quantities with an index $0$ are evaluated at the current age of the universe. Finally, the leading factor 2 ensures that both particles and anti-particles are counted towards the final abundance.

\subsection{Required coupling for the correct relic abundance}
\begin{figure}
	\centering
	\includegraphics[width=0.495\textwidth]{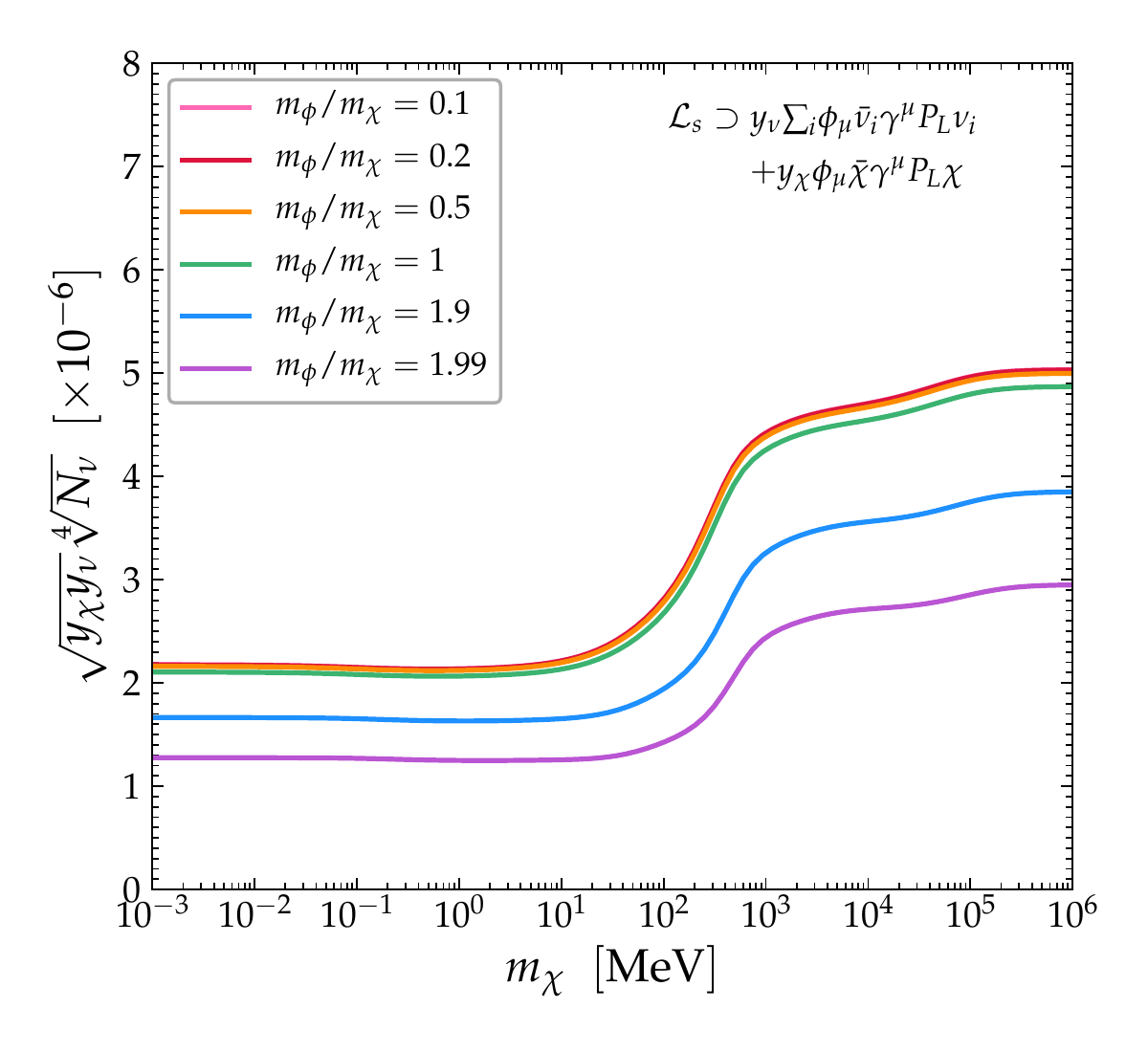}
	\includegraphics[width=0.495\textwidth]{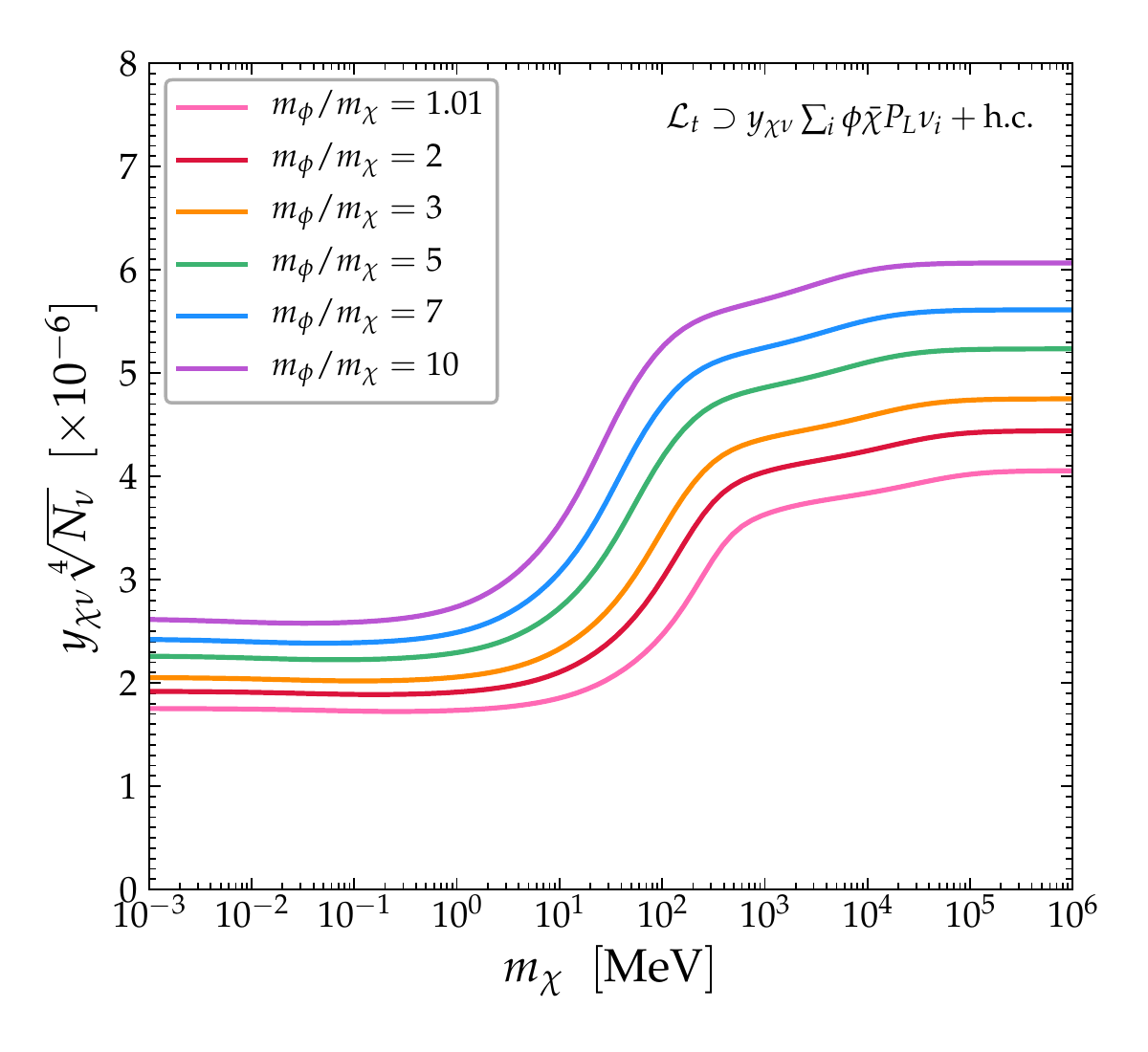}
	\hfill
	\caption{Required coupling strength $y\sqrt[4]{N_\nu}$ to obtain the correct relic abundance for different mass ratios $m_\chi/m_\psi$ in the $s$- (left) and $t$-channel (right) scenario. 
	}
	\label{fig:relic}
\end{figure}

In this section, we explicitly solve eq.~\eqref{eq:beq_n}, thereby fixing $y$ in such a way that we obtain the correct relic abundance, $\Omega_\chi h^2 = 0.12$~\cite{Planck:2018vyg}. In fig.~\ref{fig:relic}, we show the results of this calculation in the $y \sqrt[4]{N_\nu} - m_\chi$ plane for different mass ratios $m_\chi/m_\psi$ in the $s$- (left) and $t$-channel (right) scenarios. 
As we discussed in sec.~\ref{sec:Framework}, we set the decay width of the mediator to zero in the $s$-channel case since its effect is negligible.

In general, we find that the required coupling in both scenarios is approximately constant up to $\mathcal{O}(1)$ factors, i.e.~$y\sim \mathcal{O}(10^{-6})$, even for different DM masses $m_\chi$ and mass ratios $m_\phi/m_\chi$. Specifically, we find $y\cdot g_{*s}(m_\chi)^{-1/4}\simeq\text{const.}$ with $g_{*s}$ being the effective number of relativistic entropy degrees of freedom in the SM, which is in accordance with previous results in the literature \cite{Hall:2009bx}. This is because the direct dependence on $m_\chi$ cancels in the calculation of the relic abundance for freeze-in, which only leaves a mild dependence via a term $g_{*s}(T_\text{prod})$ with the production temperature $T_\text{prod}$, fulfilling $T_\text{prod}\sim m_\chi$, as most particles are produced close to threshold.
Large deviations to $y\sim \mathcal{O}(10^{-6})$ only occur in the $s$-channel scenario for $m_\phi/m_\chi \rightarrow 2$, as the resonance from eq.~\eqref{eq:matrix_s} leads to a much increased cross-section, which needs to be compensated for by a smaller coupling.
In addition, since $\Omega_\chi h^2 = \Omega_\text{DM} h^2$ holds along each colored line, we note that all points above the respective line lead to DM overproduction, while points below the respective lines lead to DM underproduction, in which case $\chi$ can still be a sub-component of DM. In general, we do not exclude the latter scenario in our analysis.

\subsection{The spectrum of DM particles}
\label{sec:evo_spectrum}
While the number density is enough to infer the relic abundance of the DM particle, for the calculation of the Lyman-$\alpha$ bounds on this scenario (cf.~section \ref{sub:lyman-alpha}) we are also interested in the actual spectrum of $\chi$. For this, we instead have to solve the Boltzmann equation at the level of the spectrum, which for $\chi$ and $\phi$ is given by ($E$ and $p$ always refer to the energy and momentum of the particle under consideration)
\begin{align}
\frac{\partial f_\chi}{\partial t} - Hp\frac{\partial f_\chi}{\partial p} &= \frac{\mathcal{A}_{\nu \bar{\nu} \rightarrow \chi \bar{\chi}} + \mathcal{D_{\phi\rightarrow\chi}}}{E} \label{eq:spectrum_chi}\\
\frac{\partial f_\phi}{\partial t} - Hp\frac{\partial f_\phi}{\partial p} &= \frac{\mathcal{A}_{\nu \bar{\nu} \rightarrow \phi \phi^*} - \mathcal{D}'_{\phi\rightarrow\chi}}{E}\eqsp.
\label{eq:spectrum_phi}
\end{align}
Here, $\mathcal{A}_{\nu \bar{\nu} \rightarrow \chi \bar{\chi}}$ ($\mathcal{A}_{\nu \bar{\nu} \rightarrow \chi \bar{\chi}}$) is the collision operator describing the production of DM (the mediator) via neutrino annihilation, and $\mathcal{D}_{\phi\rightarrow\chi}$ ($\mathcal{D}'_{\phi\rightarrow\chi}$) is the collision operator for the creation of $\chi$ (destruction of $\phi$) via the mediator decay.\footnote{It is $\int \frac{g_\chi \di^3 p}{(2\pi)^3} \frac{\mathcal{A}_{\nu \bar{\nu} \rightarrow \chi \bar{\chi}}}{E} = \mathfrak{C}_{\nu\bar{\nu}\rightarrow\chi \bar{\chi}}$ and $\int \frac{g_\chi \di^3 p}{(2\pi)^3} \frac{\mathcal{D}_{\phi\rightarrow \chi}}{E} = \int \frac{g_\chi \di^3 p}{(2\pi)^3} \frac{\mathcal{D}'_{\phi\rightarrow \chi}}{E}$.} In this this work, we specifically have $\mathcal{D}_{\phi\rightarrow\chi} = \mathcal{D}'_{\phi\rightarrow\chi} = 0$ in the $s$-channel scenario. To simplify the given set of Boltzmann equations, we explicitly assume Maxwell-Boltzmann distributions for the neutrinos in the collision operators $\mathcal{A}_{\nu \bar{\nu} \rightarrow X \bar{X}}$. This approach is somewhat justified since most particles are produced at $T_\nu \sim m_X$, and we estimated that the usage of the full quantum distributions would lead to $\lesssim 10\%$ corrections. Using this approximation, the collision operator $\mathcal{A}_{\nu \bar{\nu} \rightarrow X \bar{X}}$ can then be written as~\cite{DEramo:2020gpr}
\begin{align}
\mathcal{A}_{\nu \bar{\nu} \rightarrow X \bar{X}} & \overset{\text{MB}}{\simeq} \frac{N_\nu}{2g_X}\int \di \Pi_1 \di \Pi_2 \di \Pi_3 \;|\mathcal{M}_{\nu \bar{\nu} \rightarrow X \bar{X}}|^2 (2\pi)^4 \delta(p_1 + p_2 - p - p_3) \times e^{-E_1/T_\nu} e^{-E_2/T_\nu}\nonumber\\
& = \frac{N_\nu}{256g_X \pi^3} \frac{T_\nu e^{-E/T_\nu}}{p} \int_{4m_X^2}^\infty \di s\;\frac{e^{-E_3^-/T_\nu} - e^{-E_3^+/T_\nu}}{\sqrt{\lambda(s, m_X)}}\int_{t_-}^{t_+} \di t\; |\mathcal{M}_{\nu \bar{\nu} \rightarrow X \bar{X}}|^2\eqsp.
\end{align}
Here, $\lambda(s, m_X)=s(s-4m_X^2)$ and $\di \Pi_i = \di^3 p_i/(2\pi)^3 2E_i$, while the expressions for $E_3^\pm$ and $t_\pm$ are rather cumbersome but can be found in appendix C of~\cite{DEramo:2020gpr}. In order to solve the Boltzmann equation, let us first take a closer look at the simpler $s$-channel scenario, in which case we have $1/\tau_{\phi\rightarrow\chi}=0$ and consequently
\begin{align}
f_\chi(p, t) & \overset{s\text{-ch}}{=} \int_{t_0}^t \di t'\; \frac{\mathcal{A}^{(s)}_{\nu \bar{\nu} \rightarrow \chi \bar{\chi}}}{E}\Bigg|_{p\,\rightarrow\,pR(t)/R(t')}
\label{eq:spec_sch}
\end{align}

\begin{figure}
	\centering
	\includegraphics[width=0.495\textwidth]{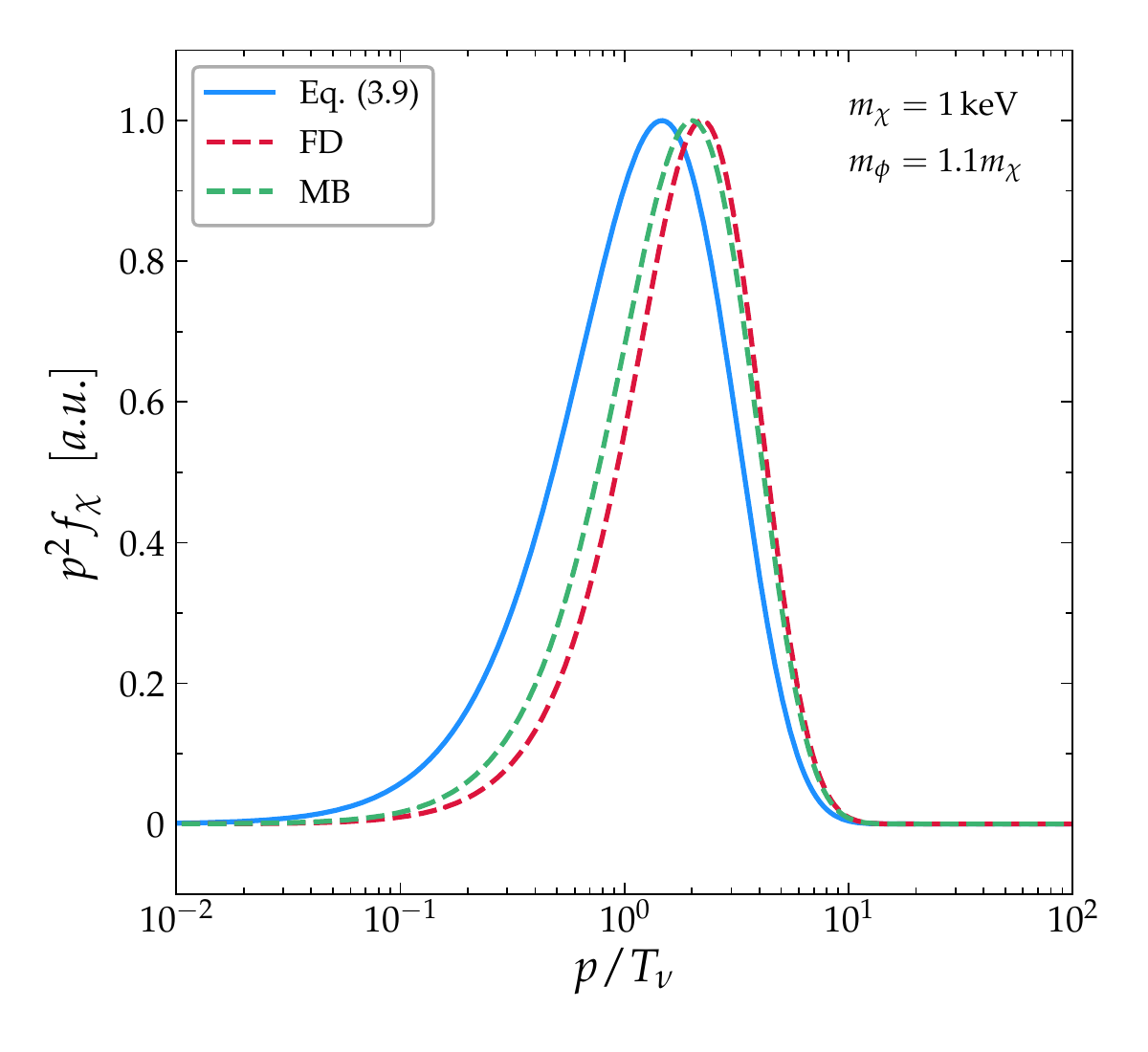}
	\includegraphics[width=0.495\textwidth]{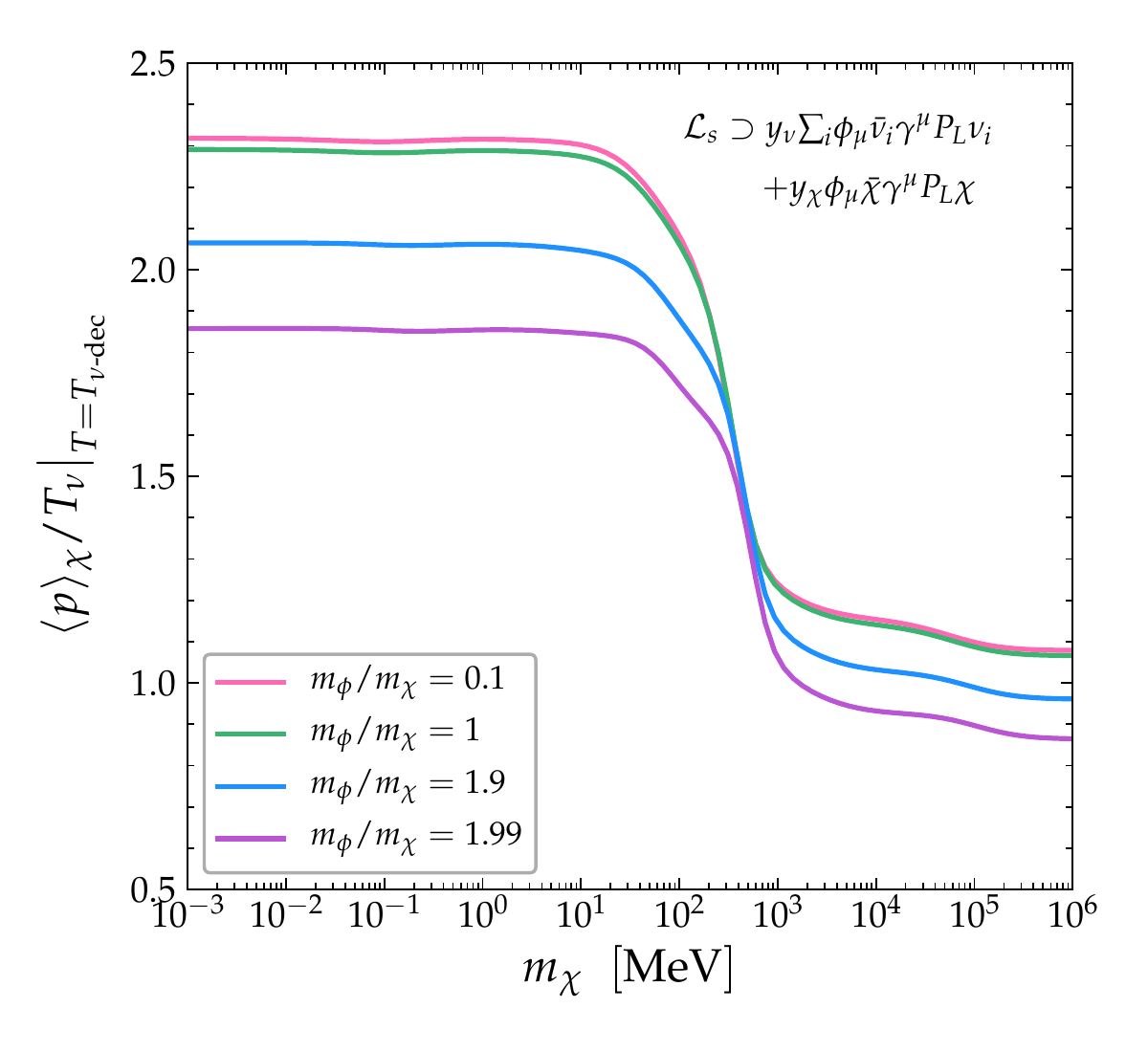}
	\caption{\emph{Left:} The phase-space distribution $p^2f_\chi$ of $\chi$ from eq.~\eqref{eq:spec_sch} for $m_\chi = 1\,\mathrm{keV}$ and $m_\phi=1.1m_\chi$ for the $s$-channel scenario (blue) in comparison to the relativistic Fermi-Dirac (red) and Maxwell-Boltzmann distribution (green). All distributions are normalized to their maximal value in order to allow for a better comparison of their shape. \emph{Right:} The average momentum over neutrino temperature $\langle p \rangle_\chi/T_\nu$ at the time of neutrino decoupling as a function of $m_\chi$ for $y=2\times10^{-6}$ and different mass splittings $m_\chi/m_\chi$ (different colors) in the $s$-channel scenario.}
	\label{fig:distortion}
\end{figure}
In the left panel of fig.~\ref{fig:distortion}, we show the resulting form of this spectrum (solid blue) for the example point $m_\chi = 1\,\mathrm{keV}$ and $m_\phi/m_\chi = 1.1$, normalized to its maximal value, meaning that the actual value of $y$ does not matter. For comparison, we also show the relativistic Fermi-Dirac (red) and Maxwell-Boltzmann (green) spectra, again normalized to their maximal values in order to allow for a better comparison of the distribution shape. As the plot shows, the actual shape of $f_\chi$ is not too different from a thermal distribution, but  the value of $p/T_\chi$ at the maximum is lower, meaning that it is slightly colder than a thermal distribution. We can illustrate this fact further by defining the average momentum $\langle p \rangle_\chi$ of $\chi$ via the relation
\begin{align}
\langle p\rangle_\chi\equiv\frac{\int p\ f_{\chi}(p)\,\di^3 p}{\int f_{\chi}(p)\,\di^3 p}\eqsp,\label{eq:ap_def}
\end{align}
which in case of a relativistic Fermi-Dirac (Maxwell-Boltzmann) distribution yields $\langle p \rangle_\chi \approx 3.15T_\nu$ ($\langle p \rangle_\chi= 3T_\nu$). Sine $f_\chi \propto y^4$ for the $s$-channel mediator, $\langle p \rangle_\chi$ is independent of $y$ in this case. For comparison, in the right panel of fig.~\ref{fig:distortion}, we show this quantity as a function of $m_\chi$ for different mass ratios $m_\phi/m_\chi$ (different colors). Note that we explicitly evaluate $\langle p \rangle_\chi$ at the neutrino decoupling temperature $T_{\nu\text{dec}}$ as $\langle p \rangle_\chi R \sim \text{const}$, meaning that the ratio $\langle p \rangle_\chi/T_\nu$ no longer changes  below $T_{\nu\text{-dec}}$.\footnote{Also this is exactly the value that we need for the Lyman-$\alpha$ constraints in sec.~\ref{sec:Constraints}.} In accordance with the results from the left panel of fig.~\ref{fig:distortion}, we find that $\langle p \rangle_\chi/T_\nu$ in our scenarios is always smaller than the corresponding value of a Fermi-Dirac or even Maxwell-Boltzmann distribution. More precisely, we find $\langle p \rangle_\chi/T_\nu \lesssim 2.32T_\nu$ for all masses and mass ratios in the $s$-channel scenario. Hence, we conclude that the final DM spectrum has indeed non-thermal features.

Performing the same calculation for the $t$-channel scenario is a little more involved, since the Boltzmann equation~\eqref{eq:spectrum_chi} for $f_\chi$ also depends on $f_\phi$. However, the respective equation~\eqref{eq:spectrum_phi} for $f_\phi$ is self-consistent, meaning that we can \textit{(1)} determine $f_\chi$ by solving eq.~\eqref{eq:spectrum_phi} and \textit{(ii)} use $f_\phi$ to calculate the DM spectrum via
\begin{align}
f_\chi(p, t) & \overset{t\text{-ch}}{=} \int_{t_0}^t \di t'\; \frac{\mathcal{A}^{(t)}_{\nu \bar{\nu} \rightarrow \chi \bar{\chi}} + \mathcal{D}^{(t)}_{\phi\rightarrow\chi}}{E}\eqsp.
\label{eq:spec_tch}
\end{align}

Overall, the calculation of $\mathcal{D}^{(t)}_{\phi\rightarrow\chi}$ and $\mathcal{D}'^{(t)}_{\phi\rightarrow\chi}$ is rather involved, especially since we cannot neglect the spin-statistical factors for the final-state neutrino. In general, we find that the results in this case are rather similar to the ones of the $s$-channel scenario. Hence, for simplicity we set $\mathcal{D}^{(t)}_{\phi\rightarrow\chi}$, in which case eq.~\eqref{eq:spec_tch} gives a lower bound for $\langle p \rangle_\chi$ and thus conservative bounds later on (cf.~sec.~\ref{sec:Constraints}). This way we find $\langle p \rangle_\chi / T_\nu \in [2.30, 2.48]$ for $m_\phi/m_\chi \in [1.01, 10]$.

\section{Cosmological constraints}
\label{sec:Constraints}
The results shown in fig.~\ref{fig:relic} are applicable as long as the final DM abundance is small compared to the abundance of neutrinos. This is true roughly until matter-radiation equality, meaning that our results can be used for a large range of masses. However, especially in the low-mass region, there exist several constraints, which limit the available parameter space. Here, the most important ones come from measurements of the Lyman-$\alpha$ forest, phase-space occupation for fermionic DM, and changes to $N_\text{eff}$, all of which will be discussed below.

\subsection{Constraints from Lyman-$\alpha$ data}
\label{sub:lyman-alpha}

Dark matter in the $\mathrm{keV}$ range constitutes warm dark matter (WDM), which can significantly suppress the matter power spectrum of density perturbations at small scales $\lesssim0.1\,\mathrm{Mpc}$.\footnote{See e.g.~fig.~1 in \cite{Kuhlen:2012ft}.} Consequently, this mass range can be constrained with recent measurements of the Lyman-$\alpha$ forest~\cite{Baur:2017stq,Irsic:2017ixq,Palanque-Delabrouille:2019iyz,Garzilli:2019qki}.

When it comes to DM (and also neutrinos), the corresponding constraints are mainly determined by the free-streaming length of the particles. Given, for example, WDM with a mass $m_\chi=100\,\mathrm{eV}$ and a thermal distribution $f_{\chi}=1/[\exp(p/T_{\nu})+1]$ identical to the one of neutrinos, the free-streaming length $\lambda_{{\rm FS}}$ is roughly $4\pi\,\mathrm{Mpc}$~\cite{Colombi:1995ze}. If instead the distribution is non-thermal, $\lambda_{{\rm FS}}$ can be estimated via the relation~\cite{Heeck:2017xbu}
\begin{equation}
\lambda_\text{FS}\simeq 4\pi\,\mathrm{Mpc} \times \left(\frac{100\,\mathrm{eV}}{m_\chi}\right)\left(\frac{\langle p\rangle_\chi}{3.15T_{\nu}}\right)\eqsp,\label{eq:lfs}
\end{equation}
with the average momentum $\langle p \rangle_\chi$ from eq.~\eqref{eq:ap_def}, which needs to be evaluated at the time of structure formation, just like $T_\nu$.

In the literature, Lyman-$\alpha$ bounds are usually reported as limits on the mass of either sterile neutrinos $\nu_s$ from the non-resonant production (NRP) mechanism~\cite{Dodelson:1993je}, or thermal relics $x$ following a Fermi-Dirac distribution~\cite{Colombi:1995ze}. In fact, using \textsc{SDSS}~\cite{SDSS:2012gam}, \textsc{XQ-100}~\cite{refId0}, \textsc{HIRES}~\cite{Vogt:1994fao}, and \textsc{MIKE}~\cite{MIKE} data as described in~\cite{Baur:2017stq}, the corresponding constraints restrict the allowed masses to $m_{\nu_s} \gtrsim 28.8\,\mathrm{keV}$ and $m_x \gtrsim 4.65\,\mathrm{keV}$, respectively. And while these bounds on the mass are rather different in both scenarios, they becomes more or less compatible when recasting them to limits on $\lambda_{{\rm FS}}$, which can then be used to derive a somewhat consistent bound on $m_\chi$ via eq.~\eqref{eq:lfs}.

In fact, for NRP sterile neutrinos $\nu_s$, production proceeds via active-sterile neutrino oscillations, leading to a distribution function of the form $f_{\nu_s} \propto 1/[\exp(p/T_\nu)+1]$ with $\langle p \rangle_{\nu_s} \simeq 3.15T_\nu$ at the time of production, $T_{\nu_s \text{-prod}} \sim 100\,\mathrm{MeV}$. However, due to entropy dilution, this quantity still changes until neutrino decoupling at $T_{\nu\text{-dec}} \approx 1.4\,\mathrm{MeV}$ by a factor $[g_{*s}(T_{\nu\text{-dec}})/g_{*s}(T_{\nu_s\text{-prod}})]^{1/3}\approx 0.85$. Plugging the so-reduced value back into eq.~\eqref{eq:lfs} we then obtain
\begin{align}
\lambda^{(\nu_s)}_\text{FS} \simeq 0.037\,\mathrm{Mpc}\times \left( \frac{28.8\,\mathrm{keV}}{m_{\nu_s}} \right)\eqsp,
\end{align}
which implies $\lambda^{(\nu_s)}_\text{FS} \lesssim 0.037\,\mathrm{Mpc}$ for the aforementioned bound $m_{\nu_s} \gtrsim 28.8\,\mathrm{keV}$.

When instead considering thermal relics $x$ with a Fermi-Dirac distribution with temperature $T_x \ll T_\nu$ 
due to entropy dilution, 
we have $\langle p \rangle_x \simeq 3.15T_\nu \times (T_x/T_\nu) \ll 3.15T_\nu$. Assuming that $x$ is a neutrino-like\footnote{Meaning that the degrees of freedom are identical.} particle that decouples when being relativistic, we can write~\cite{Colombi:1995ze}\footnote{Hence, a single neutrino flavor with a mass around $10\,\mathrm{eV}$ could account for the entire amount of DM.}
\begin{align}
\Omega_x h^2 = \left( \frac{m_x}{94\,\mathrm{eV}} \right) \left( \frac{T_x}{T_\nu} \right)^3\eqsp.
\label{eq:oh2_TT}
\end{align}
Then, by enforcing the current results from \textsc{PLANCK}~\cite{Planck:2018vyg}, $\Omega_x h^2 = 0.12$, and writing $T_x/T_\nu$ as a function of $m_x$ according to eq.~\eqref{eq:oh2_TT}, we can again use eq.~\eqref{eq:lfs} to obtain
\begin{align}
\lambda^{(x)}_\text{FS} \simeq 0.036\,\mathrm{Mpc}\times \left( \frac{4.65\,\mathrm{keV}}{m_x} \right)^{4/3}
\eqsp,
\end{align}
which yields $\lambda^{(x)}_\text{FS} \lesssim 0.036\,\mathrm{Mpc} \simeq \lambda^{(\nu_s)}_\text{FS}$ for $m_x \gtrsim 4.65\,\mathrm{keV}$.

Let us note that while these two results for $\lambda_\text{FS}$ are indeed compatible, some references report different bounds on $m_x$. For example, \cite{Palanque-Delabrouille:2019iyz} reports $m_{x}\geq5.3\,\mathrm{keV}$ and \cite{Garzilli:2019qki} claims a more conservative bound of $m_{x}\geq1.9\,\mathrm{keV}$. However, for the sake of consistency, we rather adapt the results of~\cite{Baur:2017stq} and hence use $\lambda_\text{FS} \lesssim 0.036\,\mathrm{Mpc}$ to set a bound on $m_\chi$. Nevertheless, we checked that such $\mathcal{O}(1)$ variations in the bound on $\lambda_\text{FS}$ do not significantly change our discussion. Plugging the so-obtained bound on $\lambda_\text{FS}$ back into eq.~\eqref{eq:lfs}, we finally get
\begin{align}
m_\chi \gtrsim 35\,\mathrm{keV} \times \frac{\langle p \rangle_\chi}{3.15 T_\nu}\eqsp.
\label{eq:lya-bound}
\end{align}

\begin{figure}
\centering
\includegraphics[width=0.7\textwidth]{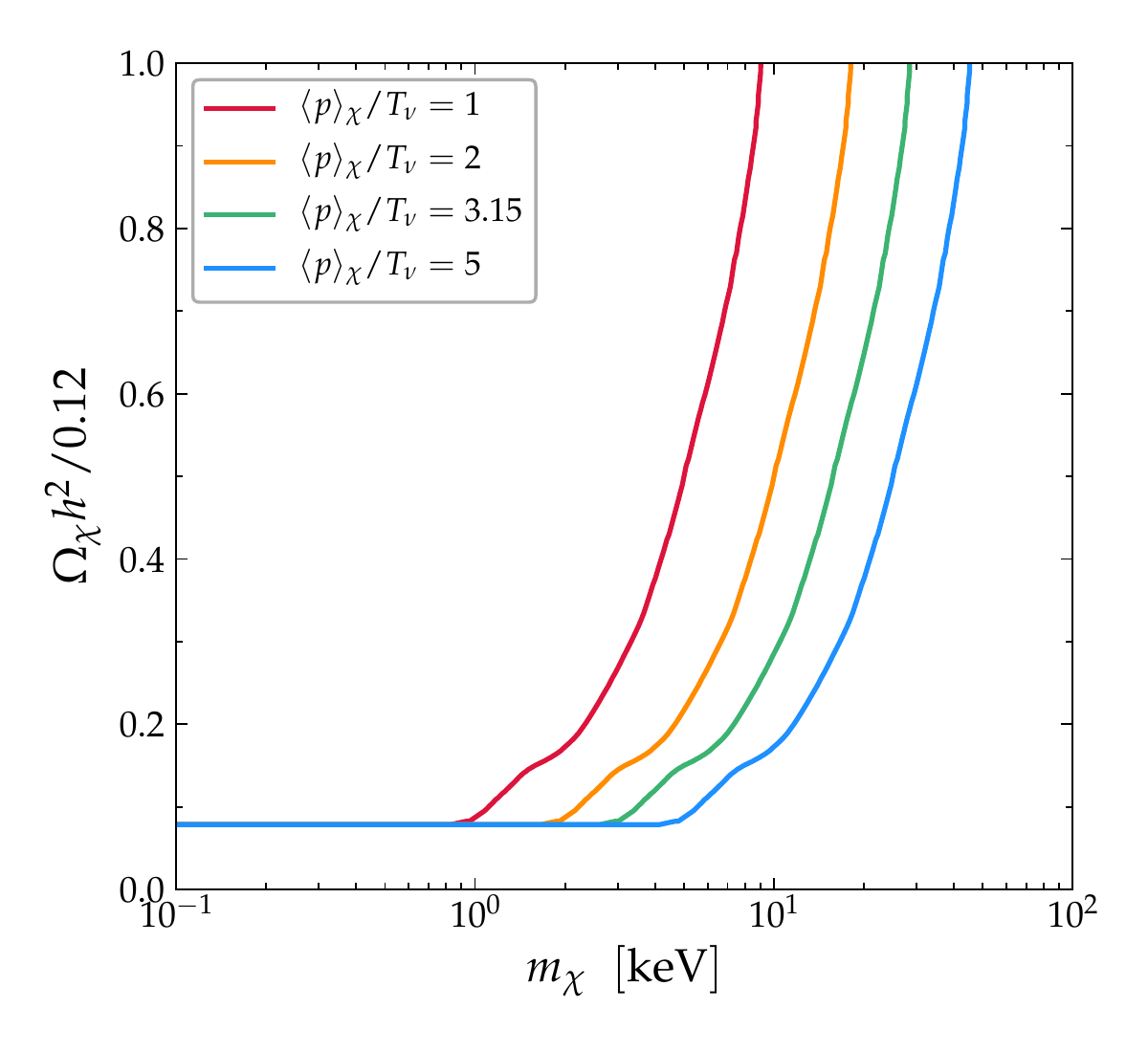}
\caption{$95\%$ C.L.~Lyman-$\alpha$ bounds from~\cite{Baur:2017stq} on the mass $m_{\chi}$ for $\Omega_{{\rm \chi}}h^{2}/0.12 \in [0, 1]$ and different values of $\langle p \rangle_\chi/T_\nu$ (different colors).}
\label{fig:lyman}
\end{figure}

Note that this bound is only applicable if $\chi$ accounts for the total DM abundance, i.e.~if $\Omega_\chi h^2 = 0.12$. However, we also want to consider scenarios in which $\chi$ is only a sub-component of DM, as motivated by some of the challenges of the $\Lambda$CDM model (cf.~\cite{Zyla:2020zbs} for a review). Updated bounds for this more general scenario can be found in~\cite{Baur:2017stq}, which we recast to a limit on $m_\chi$ for $\Omega_\chi h^2 < 0.12$ by using the $95$ C.L.~exclusion line from fig.~6 of this reference. The results of this mapping are shown in fig.~\ref{fig:lyman}.
Compared to eq.~\eqref{eq:lya-bound}, these curves lead to a slightly weaker bound for $\Omega_\chi h^2/0.12=1$, as they were computed in a full two-dimensional analysis (cf.~also the full explanation in \cite{Baur:2017stq}). Then, for a given set of parameters $(m_\chi, m_\phi)$ we can calculate $\langle p \rangle_\chi/T_\nu$ via the formalism presented in sec.~\ref{sec:evo_spectrum} and afterwards interpolate the lines in fig.~\ref{fig:lyman} to obtain the actual constraints for our specific value $\langle p \rangle_\chi/T_\nu$. This bound can then further be translated into a bound on the coupling $y$ by using
\begin{align}
y = y_\text{relic} \times \left( \frac{\Omega_\chi h^2}{0.12} \right)^{1/4}\eqsp,
\end{align}
where $y_\text{relic}$ is the required coupling to obtain $\Omega_\chi h^2 = 0.12$ (cf.~fig.~\ref{fig:relic}). This statement holds true in both of our scenarios.

\subsection{Constraints from phase-space occupation}

Since $\chi$ is considered to be fermionic DM, there exists a lower bound on the mass of $\chi$ from Pauli blocking, when assuming that $\chi$ accounts for the entire amount of DM in the Milky Way or other galaxies. In fact, given a known DM mass distribution $\rho_\chi$ in galaxies, a sufficiently small value of $m_\chi$ would imply a large number density $n_\chi = \rho_\chi/m_\chi$, possibly exceeding the limit imposed by the Fermi-Dirac statistics, as each non-relativistic $\chi$ occupies a finite spatial volume. Constraints of this type have already been derived in a series of studies~\cite{Tremaine:1979we,Boyarsky:2008ju,Angus:2009sw,DiPaolo:2017geq,Savchenko:2019qnn,Alvey:2020xsk}, and while they are generally weaker than bounds from Lyman-$\alpha$ data, they are also less model-dependent.

In order to estimate this bounds, let us note that the Fermi-Dirac distribution is always smaller than unity, while $f_\chi = 1$ corresponds to a fully degenerate state. Furthermore, DM particles in the galactic halo must have velocities below the escape velocity $v_\text{esc}$, corresponding to a maximal momentum $p_\text{max} = m_\chi v_\text{esc}$. The number density $n_\chi^{\text{cap}}$ of particles not escaping the galaxy is thus given and limited by
\begin{align}
n_\chi^{\text{cap}} = \int_0^{p_\text{max}} f_\chi \frac{4\pi p^2 \di p}{(2\pi)^3} \leq \int_0^{p_\text{max}} \frac{4\pi p^2 \di p}{(2\pi)^3} = \frac{p_\text{max}^3}{6\pi^2}\eqsp.
\end{align}

Then by using the local DM density $\rho_{{\rm DM}}\approx0.4\,{\rm GeV}/{\rm cm}^{3}$~\cite{Read:2014qva} with $\rho_{{\rm DM}}=2n_{\chi}^{\text{cap}}m_{\chi}$ and the escape velocity $v_{{\rm esc}}\approx550\,{\rm km}/{\rm sec}$, we find that only masses with $m_{\chi}\gtrsim11\,\mathrm{eV}$ are allowed. Using DM densities from other locations in the Galaxy with their respective escape velocities would lead to similar results, but also introduce a dependence on the halo profile.

\subsection{Constraints from $\Delta N_{{\rm eff}}$}

Finally, the production of DM might also change the effective number $N_\text{eff}$ of neutrinos if this process concludes after neutrino decoupling. However, the most recent \textsc{PLANCK} measurements find $|\Delta N_\text{eff}| < 0.396$~\cite{Planck:2018vyg} at $2\sigma$, meaning that this constraint can only become relevant when a substantial amount of DM is converted into neutrinos, in which case eq.~\eqref{eq:beq_n} is no longer valid. Instead, it becomes necessary to solve the full set of coupled Boltzmann equations for the neutrinos, the DM particle, and potentially even the mediator. In this work, we will not go this route, but instead make a suitable estimate in order to approximate $ \Delta N_\text{eff}$. To this end, let us note that the production process $\nu \bar{\nu} \rightarrow \chi\bar{\chi}$ conserves the combined number of neutrinos in both scenarios. In case of $t$-channel annihilation, however, there exists the additional process $\nu \bar{\nu} \rightarrow \phi\phi^*$ as well as subsequent decay $\phi\rightarrow \chi \bar{\nu}$, which does not produce any new neutrinos. Hence, if $n_\chi$ is the number of DM particles not originating from the $\phi$ decay (analogously to eq.~\eqref{eq:def_ns}), we have $n_\nu = n_\nu^\text{st} - n_\chi$ with $n_\nu^\text{st}$ being the number density od neutrinos in the standard thermal history. If we then further assume that the production process does not significantly change the neutrino spectrum, meaning that it remains approximately Fermi-Dirac, we also have $\rho_\nu \simeq k n_\nu T_\nu = \rho_\nu^\text{st} - k n_\chi T_\nu$ with $k = 7\pi^4/(180\zeta(3)) \approx 3.15$. Using this expression for the calculation of the effective number of neutrinos, we thus obtain
\begin{align}
N_\text{eff}^0 + \Delta N_\text{eff} = \frac{\rho_{\nu, \text{rec}}}{\rho_\text{ref}} \simeq \frac{\rho^\text{st}_{\nu,\text{rec}} - kn_{\chi, \text{rec}} T_{\nu, \text{rec}}}{\rho_\text{ref}} = N_\text{eff}^0 -\frac{k n_{\chi,\text{rec}} T_{\nu, \text{rec}}}{\rho_\text{ref}}
\end{align}
with $N_\text{eff}^0 = 3.046$ and $\rho_\text{ref} = 2\times(7/8)(\pi^2/30)T_\text{rec}^4(4/11)^{4/3}$, all evaluated at the time of recombination $t_\text{rec}$ as indicated by the index ``rec''. Consequently,
\begin{align}
\Delta N_\text{eff} \simeq -3.15 \times \frac{n_{\chi, \text{rec}} T_{\nu,\text{rec}}}{\rho_\text{ref}}\eqsp.
\end{align}
Hence, for a given set of parameters $(m_\chi, m_\phi, y)$, this relation can be used to approximate $\Delta N_\text{eff}$ from the solution $n_\chi$ of eq.~\eqref{eq:beq_n}, i.e.~without the neutrino-number preserving mediator decay. We have to keep in mind, however, that at some point the inverse reaction $\chi \bar{\chi} \rightarrow \nu\bar{\nu}$ becomes important, which transforms some of the DM particles back into neutrinos. Hence, by taking $n_\chi$ from eq.~\eqref{eq:beq_n} we might use a number density that is a little too large, thus making the bound too aggressive. However, we will later see that this bound is still weaker than the others and therefore does not contribute to the overall exclusion limit.

Nevertheless, this statement also implies that $\Delta N_\text{eff}$ can be used to quantify whether our calculation is still correct. If $\Delta N_\text{eff} \sim 0.3$, roughly $10\%$ of neutrinos are converted into DM, meaning that our general approximations (back reaction is negligible and neutrino spectrum remains Fermi-Dirac) are no longer appropriate. We will utilize this statement in more detail in sec.~\ref{sec:nu_signal}.

\section{Prospects for a neutrino signal from DM annihilation}
\label{sec:nu_signal}

An interesting feature of our model is the potential existence of observable neutrino signals fom DM annihilation, i.e. via the process $\chi\bar{\chi}\rightarrow\nu\bar{\nu}$, which can happen in the galactic center due to the large DM number density. In the next section, we will discuss the detection prospects of such a signal.

\subsection{Neutrino flux from the galactic center}

In general, the neutrino flux $\Phi_\nu$ from DM annihilation in the galactic center can be calculated from the relation~\cite{Hooper:2018kfv}
\begin{equation}
\Phi_{\nu}=\frac{1}{2}\frac{\langle\sigma v\rangle_{\chi\bar{\chi}\rightarrow\nu\bar{\nu}}}{4\pi m_{\chi}^{2}}J\eqsp,\label{eq:nu_flux}
\end{equation}
where $J$ is often referred to as the {\it $J$ factor} which will be explained later, and $\langle\sigma v\rangle_{\chi\bar{\chi}\rightarrow\nu\bar{\nu}}$ is the thermally averaged annihilation cross section.

In the special case of non-relativistic $s$-wave annihilation,\footnote{For all interactions considered in this work, this statement is always true. This is because, if the mediators are integrated out -- up to Fierz transformations -- the effective interactions are vector-like, which always leads to $s$-wave annihilation. See e.g.~the discussion in \cite{Hambye:2021xvd}.}  the latter quantity is given by
\begin{align}
\langle\sigma v\rangle_{\chi\bar{\chi}\rightarrow\nu\bar{\nu}} \simeq \frac{|{\cal M}|^{2}}{32\pi m_{\chi}^{2}}\eqsp,\label{eq:sv2_flux}
\end{align}
with the matrix elements from eq.~\eqref{eq:matrix_s} and~\eqref{eq:matrix_t} for $s$- and $t$- channel annihilation, respectively. However, since this process happens when the DM particles are almost at rest, we can set $s\simeq 4m_\chi^2$ and $t\simeq u \simeq -m_\chi^2$. Eq.~\eqref{eq:sv2_flux} holds true as long as Sommerfeld enhancement~\cite{Sommerfeld} is absent or insignificant, i.e.~for $y_{\chi}^{2}m_{\chi}/m_{\phi}\ll4\pi$. For the $s$-channel scenario, however, the vector boson induces an attractive potential between the slow-moving $\chi$ and $\bar{\chi}$ particles, which can lead to a significant enhancement of their annihilation cross-section. We take this effect into account by replacing $\langle \sigma v\rangle_{\chi\bar{\chi}\rightarrow\nu\bar{\nu}} \rightarrow \bar{S} \langle \sigma v\rangle_{\chi\bar{\chi}\rightarrow\nu\bar{\nu}}$ with the velocity-averaged Sommerfeld factor $\bar{S}$. More details on this can be found in appendix~\ref{sec:Sommerfeld}.

Moreover, the $J$ factor is determined by the DM distribution in the galaxy and is given by~\cite{Hooper:2018kfv}
\begin{align}
J\equiv\int_{\Delta\Omega}\di\Omega\int_{0}^{\infty}\di l\,\rho_{\chi}^{2}(l,\ \Omega)\eqsp.\label{eq:jfactor}
\end{align}
In this expression, the integral $\int\di\Omega$ is evaluated over a solid
angle $\Delta\Omega$ \footnote{For neutrino detectors we integrate over the full unit sphere.} and $\int\di l$ is evaluated along the \emph{line-of-sight}. Regarding the  energy density of DM in the galaxy $\rho_\chi$, we employ the frequently used Navarro-Frenk-White (NFW) halo profile~\cite{Navarro:1995iw,Navarro:1996gj}, which is given by
\begin{equation}
\rho_{\chi}(r)=\frac{\rho_{\astrosun}}{\left(\frac{r}{R}\right)\left(1+\frac{r}{R}\right)^{2}}\eqsp.\label{eq:nfw}
\end{equation}
Here, $R \approx 20\,\mathrm{kpc}$ is the scale radius of the galaxy, $\rho_{\astrosun} \approx 0.4\,\mathrm{GeV}/\mathrm{cm}^3$ is the local matter density of the solar system, and $r$ is the distance to the galactic center. Moreover, the quantity $r$ is related to $l$ and $\Omega$ via
\begin{equation}
r=\sqrt{r_{\astrosun}^{2}-2r_{\astrosun}l\cos\theta+l^{2}}\eqsp,
\end{equation}
with the distance $r_{\astrosun}$ between the sun and the galactic center, and the angle $\theta$ along the line of sight, i.e. $\theta=0$ encodes the direction directly in the line-of-sight. Hence, $r\rightarrow r(l, \Omega)$ and consequently $\rho_\chi(r) \rightarrow \rho_\chi(l, \Omega)$. Plugging this expression for $\rho_\chi(l, \Omega)$ back into eq.~\eqref{eq:jfactor}, we find
\begin{align}
J\approx2.1\times10^{23}\,\mathrm{GeV}^{2}/\mathrm{cm}^{5}\eqsp.\label{eq:japprox}
\end{align}
Using this value to evaluate eq.~\eqref{eq:nu_flux}, we finally obtain
\begin{align}
\Phi_\nu = \left(\frac{y}{10^{-6}}\right)^{4}\left(\frac{100\,{\rm eV}}{m_{{\rm eff}}}\right)^{4} \times \begin{cases}
3.9\times10^{7}\ \text{cm}^{-2}\sec^{-1}&\qquad t\text{-channel}\\
1.6\times10^{8}\ \text{cm}^{-2}\sec^{-1} \times \bar{S}&\qquad s\text{-channel}
\end{cases}
\end{align}
with
\begin{align}
m_\text{eff}^{4}\equiv\begin{cases}
(m_{\chi}^{2}+m_{\phi}^{2})^{2}&\qquad t\text{-channel}\\
(4m_{\chi}^{2}-m_{\phi}^{2})^{2}+(m_{\phi}\Gamma_{\phi}^{(s)})^2&\qquad s\text{-channel}
\end{cases}\eqsp.
\end{align}

\begin{figure}[t]
	\centering
	\includegraphics[width=0.7\textwidth]{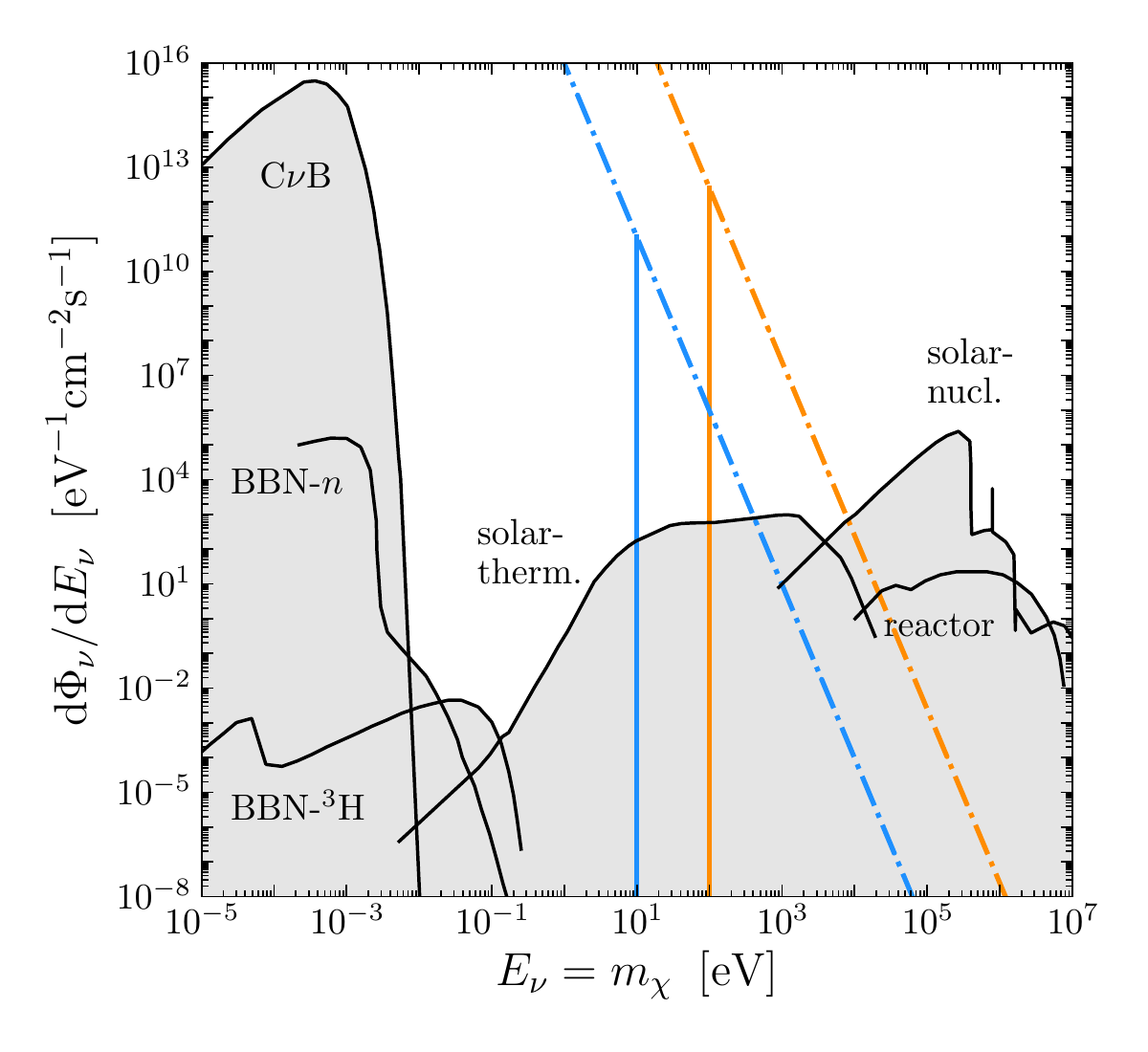}
	\caption{The neutrino flux from DM annihilation $\chi\bar{\chi}\rightarrow\nu\bar{\nu}$ in the galactic center for $m_\chi=10\,\mathrm{eV}$, $m_\phi=1.01m_\chi$ in the $t$-channel scenario (blue line), as well as $m_\chi=100\,\mathrm{eV}$, $m_\phi\simeq 0.6m_\chi$ in the $s$-channel scenario (orange line), compared to the relevant backgrounds from~\cite{Vitagliano:2019yzm} (black lines). In addition, the dotted lines show the change in height when varying the value of $m_{\chi}$, corresponding to a variation of $m_\text{eff}$.}
	\label{fig:neutrino-signal}
\end{figure}

Since the annihilating DM particles can be considered at rest, the neutrino flux is approximately monochromatic with energy $E_\nu = m_\chi$. In fig.~\ref{fig:neutrino-signal} we show the respective signal for the couplings $y=10^{-6}$, $y_\chi=\sqrt{4\pi}$, as well as mass combinations $m_\chi = 10\,\mathrm{eV}$, $m_\phi/m_\chi = 1.01$ ($t$-channel, blue line) and $m_\chi = 100\,\mathrm{eV}$, $m_\phi/m_\chi \simeq 0.6$ ($s$-channel, orange line).\footnote{In the $s$-channel scenario, the Sommerfeld enhancement factor $\bar{S}$ is quite sensitive to the mediator mass $m_{\phi}$. For illustration, we take the first peak value, $m_{\phi}=3m_{\chi}y_{\chi}^{2}/(2\pi^{3})\simeq 0.6 m_\chi$, and compute the enhancement from eq.~\eqref{eq:m-91} with $n=1$.} These combinations of masses explicitly correspond to $m_\text{eff}^4 \simeq 4.08m_\chi^4$ and $m_\text{eff} \simeq 13.25m_\chi^4$; however, we also vary $m_{{\rm eff}}\in[1\,{\rm eV},\ 1\,{\rm MeV}]$ in order to quantify the actual mass dependence (dashed orange/blue lines). For comparison, we also show the relevant neutrino backgrounds from~\cite{Vitagliano:2019yzm} (black lines), i.e.~the expected cosmic neutrino background (C$\nu$B) with $m_\nu = 0$, neutrinos from neutron (BBN-$n$) and tritium (BBN-${}^3$H) decays during BBN, solar neutrinos from nuclear fusion reactions (solar-nucl.) and thermal radiation (solar-therm.), as well as reactor neutrinos.
Overall, we find that the expected signal can indeed be much larger than all of the relevant backgrounds, which implies that such a process could in principle be observed in a neutrino detector with a sufficiently low detection threshold, like those employing coherent elastic neutrino scattering or beta decays (cf.~sec.~\ref{sub:ptolemy}).

Finally, let us note that the monochromatic neutrino flux in fig.~\ref{fig:neutrino-signal} would usually appear as a $\delta$ distribution. However, in a more realistic scenario, i.e.~when taking the small but non-vanishing velocity of the DM particles into account, we instead obtain a non-vanishing width of order $10^{-3}E_\nu$, which is too small to be resolved by most modern neutrino detectors. Hence, for practical reasons, we assume an energy resolution of $\Delta E_\nu/E_\nu = 10\%$ and plot $\Phi_\nu/\Delta E_\nu$ instead of $\Phi_\nu\delta(E_\nu-m_\chi)$ in fig.~\ref{fig:neutrino-signal}.

\subsection{Detection prospects with \textsc{PTOLEMY}}
\label{sub:ptolemy}

As originally proposed in~\cite{Weinberg:1962zza}, ultralow-energy neutrinos can potentially be detected via their capture by nuclei that are prone to beta decays. For example, the decay channel ${\rm ^{3}H}\rightarrow{\rm ^{3}He}\,e^{-}\bar{\nu}_{e}$ features a low $Q$ value determined by $m_{{\rm ^{3}H}}-m_{{\rm ^{3}He}}-m_{e}\approx18.6\,\mathrm{keV}$ which implies that the related capture process
\begin{equation}
\nu_{e}+{\rm ^{3}H}\rightarrow{\rm ^{3}He}+e^{-}\label{eq:h3_cap}
\end{equation}
can in principle be used to detect electron neutrinos of arbitrarily low energies, as the corresponding signal would be an electron with energy $E_e = Q + E_\nu$, i.e.~slightly above the $Q$ value due to the absorption of the neutrino.

Currently, the required technology to detect the C$\nu$B based on this process is developed for the \textsc{PTOLEMY} experiment~\cite{PTOLEMY:2018jst,PTOLEMY:2019hkd}. In fact, the final design would feature a 100 gram tritium target capable of capturing 4 (8) C$\nu$B neutrinos per year, assuming that neutrinos are Dirac (Majorana) particles. However, implementing such detection mechanism is rather challenging, as it requires measuring the kinetic energy of the emitted electron with a high precision. For example, to distinguish an actual capture event from the spontaneous decay of a tritium atom, a sub-$\mathrm{eV}$ energy resolution is required. This has been shown to be possible only recently by the \textsc{PTOLEMY} collaboration by using a novel technology that deposits tritium nuclei onto a graphene substrate. 

However, unlike C$\nu$B detection, neutrinos from DM annihilation can have much higher energies, meaning that their detection is significantly less challenging. In this context, the most important factor is the capture rate $\Gamma_\text{cap}$, which is given by $\Gamma_\text{cap}=N_3 \sigma_\text{cap} \Phi_\nu$ with the number $N_3$ of tritium atoms and the capture cross section $\sigma_\text{cap}$. The latter quantity is approximately constant for $E_\nu < Q = 18.6\,\mathrm{keV}$ and given by~\cite{Cocco:2007za,Long:2014zva}
\begin{align}
\sigma_\text{cap}=7.668\times10^{-45}\,\mathrm{cm}^{2}\eqsp.\label{eq:xcap}
\end{align}
Given that 100 grams of tritium contain $100/3$ moles of atoms, we have $N_3~\approx~2~\times~10^{25}$ and consequently\footnote{For C$\nu$B neutrinos, the flux is $\Phi_{\nu}\approx1.7\times10^{12}\ {\rm cm}^{-2}\sec^{-1}$and consequently $\Gamma_\text{cap} \approx 8\ {\rm events}/\text{year}$.}
\begin{align}
\Gamma_{\text{cap}} =1\,\mathrm{event}/\mathrm{year}\times\left(\frac{M_3}{100\,\mathrm{g}}\right)\times\left(\frac{\Phi_{\nu}}{2\times10^{11}\,\mathrm{cm}^{-2}\mathrm{s}^{-1}}\right)
\label{eq:Gcap}
\end{align}
with the fiducial mass $M_3$ of the tritium target. By inserting the flux from eq.~\eqref{eq:nu_flux} into this expression, we can then calculate the expected number of events that can be detected by \textsc{PTOLEMY}. In fact, since the neutrino energy is much higher than the one of the C$\nu$B, detection of such neutrinos  is not limited  by the sub-{\rm eV} energy resolution. Without this limitation, the fiducial mass can be expanded to a much larger scale. In addition, since the half-life of tritium is about 12 years, one can also consider a longer exposure time. Therefore, in  addition to \textsc{PTOLEMY} we also consider another experimental configuration with $M_{3}=10\,\mathrm{kg}$ and a 10 year effective exposure time. 

\begin{figure}[t]
	\centering
	\includegraphics[width=0.495\textwidth]{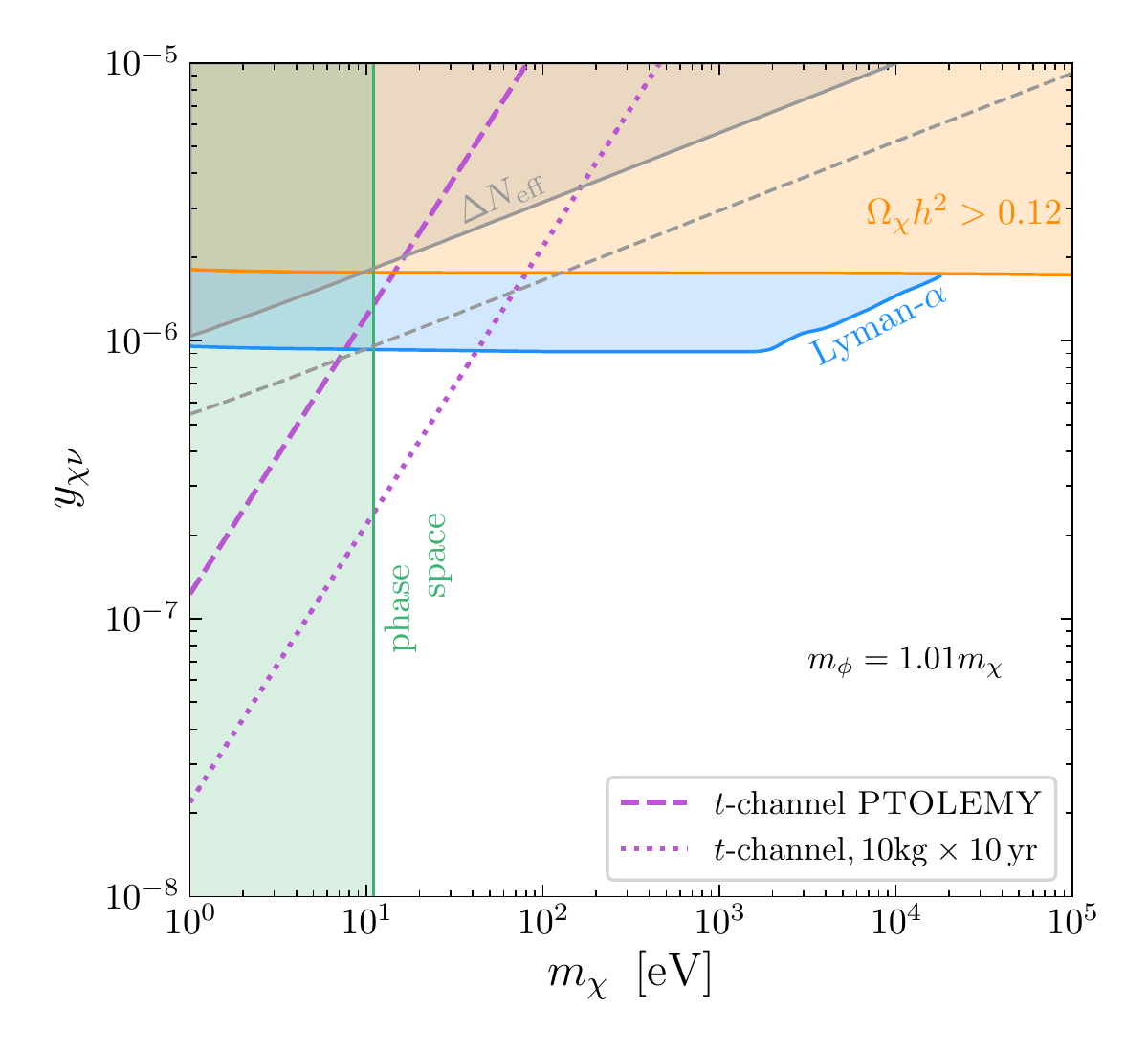}
	\includegraphics[width=0.495\textwidth]{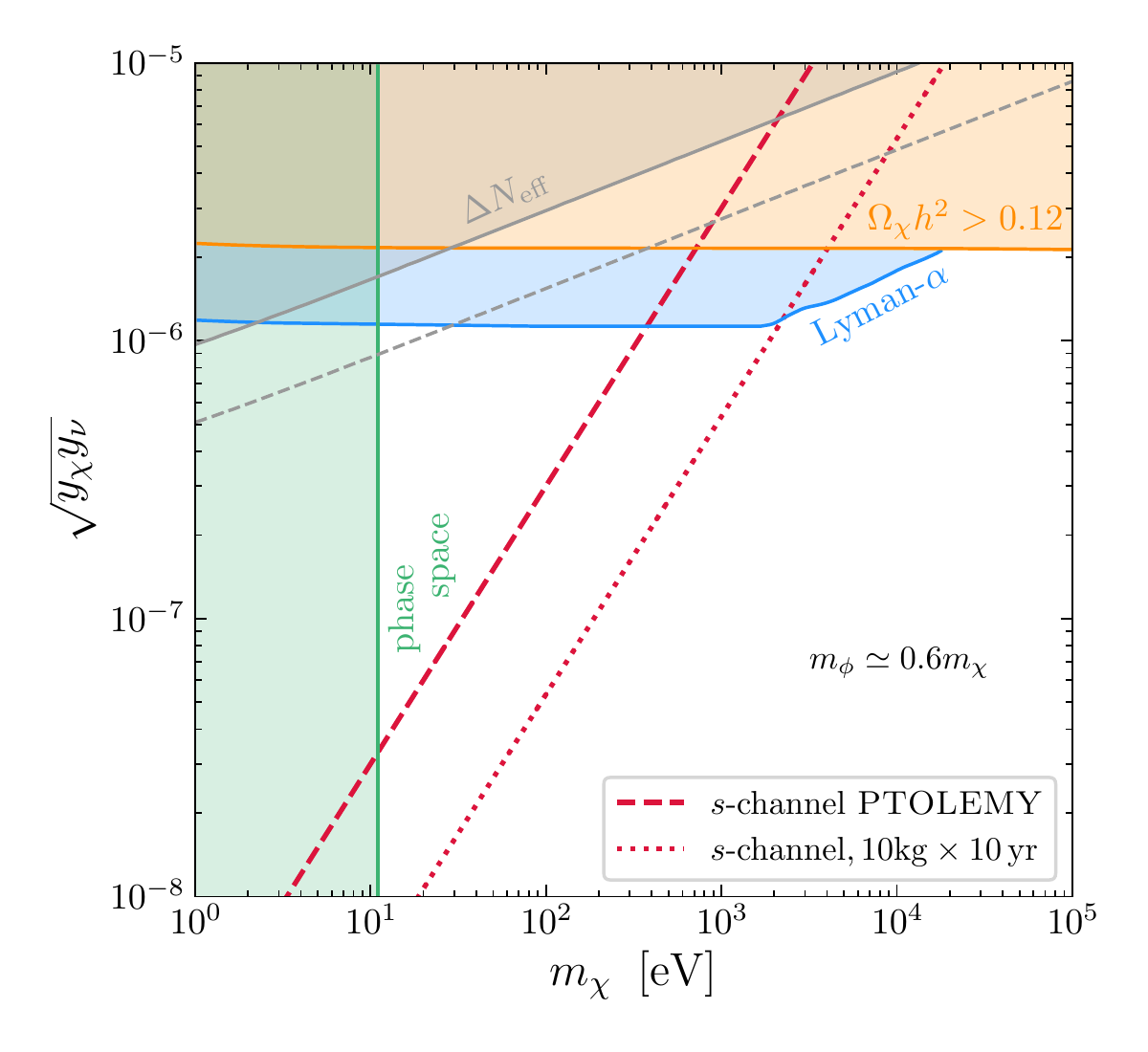}
	\caption{Prospects for \textsc{PTOLEMY}-like experiments to probe neutrino signals from DM annihilation in the $t$- (purple, left panel) and $s$-channel (red, right panel) scenario. For comparison, we also show constraints from Lyman-$\alpha$ data (blue), phase-space occupation (green), and $\Delta N_{{\rm eff}}$ (black), and DM overproduction (orange). In addition to the default configuration with $M_3 = 100\,\mathrm{g}$ (dashed), we also show results for $M_3 = 10\,\mathrm{kg}$ with a prolonged exposure time of 10 years instead of 1 year (dotted).}
	\label{fig:all}
\end{figure}
In fig.~\ref{fig:all}, we show the prospects of these two experiments for $m_\phi=1.01m_\chi$ ($t$-channel, left panel) and $m_\phi\simeq 0.6m_\chi$ ($s$-channel), i.e.~for the same mass splittings as before, in the $y-m_\chi$ parameter plane. In addition to the contours with $\Gamma_\text{cap}=1\,\mathrm{event}/\mathrm{year}$ for \textsc{PTOLEMY} (dashed purple/red) and our fictitious detector with $10\,\mathrm{kg}$ tritium and $10\,\mathrm{yr}$ of exposure time (dotted purple/red), we also show for comparison the different constraints from sec.~\ref{sec:Constraints}. These include bounds from Lyman-$\alpha$ data (blue), phase-space occupation (green), $\Delta N_\text{eff}$ (black), as well as DM overproduction (orange, cf.~fig.~\ref{fig:relic}).\footnote{Note that the line goes slightly up towards very small masses. This is because, the freeze-in happens very close to matter-radiation equality in this case, which leads to an increase in the effective number of entropy degrees of freedom.}  Hence, for points below the orange line, $\chi$ cannot account for all of DM, and we instead assume that the remaining abundance is accounted for by standard cold DM. Also, in our analysis, we explicitly set $N_\nu=1$ unlike before, as \textsc{PTOLEMY} is only sensitive to electron neutrinos.

First of all, we find that the constraints in both scenarios are rather similar. This is because, for the given mass splittings, the couplings that are required to obtain the correct relic abundance are $y\approx1.73\times10^{-6}$ and $y\approx 2.13\times 10^{-6}$ for the $t$- and $s$-channel scenario, respectively. Now, since both couplings are very close and the production processes are rather similar in both scenarios, quantities like $n_\chi$ or $\langle p \rangle_\chi/T_\nu$ will also not vary much, which ultimately leads to almost identical limits. Note that this statement will remain true for other mass splittings, as the required coupling, in both scenarios, only has a mild dependence on this parameter (cf.~fig.~\ref{fig:relic}). Hence, the following discussion will qualitatively also remain true for other mass splittings. Only for $m_\phi/m_\chi > 2$ in the $s$-channel scenario -- which we do not consider -- the required coupling would become rather small, thus leading to an exclusion of most of the parameter space. Finally, let us note that we also show the region with $\Delta N_{\text{eff}}=3\%$ (dashed gray), as above this curve it becomes questionable whether the assumptions that we made for our calculation are still justified. However, we find that this affects only a very small part of the unconstrained parameter space.

Overall, we find that in the $s$-channel scenario, there remains a lot of parameter space that can potentially be probed by \textsc{PTOLEMY}. However, these regions do not include the case where $\chi$ makes up all of DM. For this it would be necessary to go even beyond the hypothetical detector with $10\,\mathrm{kg}$ of tritium and an exposure time of $10\,\mathrm{years}$, as $\Omega_\chi h^2 = 0.12$ requires roughly $m_\chi \gtrsim 20\,\mathrm{keV}$ due to the Lyman-$\alpha$ constraint.
In comparison, the prospects for the $t$-channel scenario are slightly worse, which is mainly due to the absence of Sommerfeld enhancement. Here, the sensitivity curve of \textsc{PTOLEMY} does not exceed current backgrounds. However, further scaled-up detectors with $10\,\mathrm{kg}$ tritium  might still be able to detect a signal for $m_\chi \sim 10-50\,\mathrm{eV}$.

\section{Conclusions}
\label{sec:Conclusion}
In the presence of interactions between a DM particle $\chi$ and neutrinos, one possible way of producing $\chi$ is via freeze-in from the neutrino bath, e.g.~via reactions of the form $\nu\bar{\nu}\rightarrow \chi\bar{\chi}$. In this work, we considered two distinct scenarios, in which the relevant annihilation reactions are either mediated by a $t$-channel scalar or an $s$-channel vector boson (see figs.~\ref{fig:diagrams_1} and \ref{fig:diagrams_2}). Within these scenarios, we then calculated the DM relic abundance by solving the full Boltzmann equation at the level of the number density and found that in order to obtain the observed abundance of $\Omega_{\rm DM}h^2=0.12$, the required couplings must be $\sim 10^{-6}$, almost independent of the actual mass of $\chi$ (cf.~fig.~\ref{fig:relic}).
Additionally, by solving the Boltzmann equation at the level of the spectrum, we found that $\chi$ does not follow a thermal distribution in this scenario and generally has a lower mean momentum than a Fermi-Dirac distribution (cf.~fig.~\ref{fig:distortion}). More specifically, we found $\langle p \rangle_\chi/T_\nu\sim2$, which is smaller than the value obtained from a Fermi-Dirac distribution, i.e.~$\langle p \rangle_\nu/T_\nu \lesssim 3.15$. Hence, DM that is produced via freeze-in from neutrinos is usually colder than the neutrinos themselves. Using the solution of the Boltzmann equation, we then further studied possible constraints that need to be imposed on our scenario, including bounds from Lyman-$\alpha$ data, phase-space occupation, and $\Delta N_\mathrm{eff}$. We then compared these constraints with the potential signal from DM annihilation to neutrinos in today's galactic center. In general, we found that the \textsc{PTOLEMY} experiments, which aims at the detection of the C$\nu$B, can potentially be sensitive to such a signal. In fact, within the proposed sensitivity, a considerably large part of the $s$-channel parameter space could potentially be explored. In addition, it might also be possible to increase the detection prospects even further by building more specific experiments, in which case it would also be feasible to probe the $t$-channel scenario.

\newpage
\appendix

\section{Matrix elements, cross sections and decay widths}
\label{sec:M2}
In this appendix, we detail our calculation of the relevant matrix elements and further derive the corresponding cross sections and decay widths.

\subsection{Scattering}
For the $t$- and $s$-channel diagrams, the squared amplitudes for the process  $\nu\bar{\nu}\rightarrow\chi\bar{\chi}$ read
\begin{align}
|{\cal M}^{(t)}_{\nu\bar{\nu}\rightarrow\chi\bar{\chi}}|^{2} & =\sum_{{\rm spins}}y_{\chi\nu}^{4}|\bar{u}_{3}P_{L}u_{1}\frac{1}{t-m_{\phi}^{2}}\bar{v}_{2}P_{R}v_{4}|^{2}\eqsp,\label{eq:m-38}\\
|{\cal M}^{(s)}_{\nu\bar{\nu}\rightarrow\chi\bar{\chi}}|^{2} & =\sum_{{\rm spins}}y_{\chi}^{2}y_{\nu}^{2}|\bar{u}_{3}\gamma_{\mu}P_{L}v_{4}\frac{1}{s-m_{\phi}^{2}+im_{\phi}\Gamma_{\phi}}\bar{v}_{2}\gamma_{\mu}P_{L}u_{1}|^{2}\label{eq:m-39}\eqsp.
\end{align}
Here, $u_{\cdot}$ and $v_{\cdot}$ denote the fermion and anti-fermion spinors while the sum goes over all particle spins. Note that there is no averaging factor, since all initial-state particles are purely left-handed.
Then, by using standard trace technology as well as {\tt Package-X}~\cite{Patel:2015tea} to compute the resulting traces, we obtain
\begin{align}
|{\cal M}^{(t)}_{\nu\bar{\nu}\rightarrow\chi\bar{\chi}}|^{2} & =y_{\chi\nu}^{4}\left(\frac{1}{t-m_{\phi}^{2}}\right)^{2}{\rm tr}\left[u_{3}\bar{u}_{3}P_{L}u_{1}\bar{u}_{1}P_{R}\right]{\rm tr}\left[v_{2}\bar{v}_{2}P_{R}v_{4}\bar{v}_{4}P_{L}\right]\nonumber \\
&
=y_{\chi\nu}^{4}\left(\frac{1}{t-m_{\phi}^{2}}\right)^{2}(2p_{1}\cdot p_{3})(2p_{2}\cdot p_{4}) =y_{\chi\nu}^{4}\left(\frac{t-m_{\chi}^{2}}{t-m_{\phi}^{2}}\right)^{2}\label{eq:m-40}
\end{align}
as well as
\begin{align}
|{\cal M}^{(s)}_{\nu\bar{\nu}\rightarrow\chi\bar{\chi}}|^{2} & =\frac{y_{\chi}^{2}y_{\nu}^{2}}{(s-m_{\phi}^{2})^{2}+m_{\phi}^{2}\Gamma_{\phi}^{2}}{\rm tr}\left[u_{3}\bar{u}_{3}\gamma_{\mu}P_{L}v_{4}\bar{v}_{4}\gamma_{\nu}P_{L}\right]{\rm tr}\left[v_{2}\bar{v}_{2}\gamma^{\mu}P_{L}u_{1}\bar{u}_{1}\gamma^{\nu}P_{L}\right]\nonumber \\
 & =\frac{16y_{\chi}^{2}y_{\nu}^{2}}{(s-m_{\phi}^{2})^{2}+m_{\phi}^{2}\Gamma_{\phi}^{2}}(p_{1}\cdot p_{4})(p_{2}\cdot p_{3}) =4y_{\chi}^{2}y_{\nu}^{2}\frac{\left(u-m_{\chi}^{2}\right)^{2}}{\left(s-m_{\phi}^{2}\right)^{2}+m_{\phi}^{2}\Gamma_{\phi}^{2}}\eqsp.\label{eq:m-41}
\end{align}
Additionally, the $t$-channel diagram for the process $\nu\bar{\nu}\rightarrow\phi\phi^{*}$ in fig.~\ref{fig:diagrams_1} yields
\begin{align}
|{\cal M}_{\nu\bar{\nu}\rightarrow\phi\phi^{*}}^{(t)}|^{2} & =\sum_{{\rm spins}}y_{\chi\nu}^{4}|\bar{v}_{2}P_{R}\frac{q_\mu\gamma^\mu}{q^{2}-m_{\chi}^{2}}P_{L}u_{1}|^{2}\nonumber \\
 & =y_{\chi\nu}^{4}\left(\frac{1}{t-m_{\chi}^{2}}\right)^{2}{\rm tr}\left[v_{2}\bar{v}_{2}P_{R}q_\mu\gamma^\mu P_{L}u_{1}\bar{u}_{1}P_{R}q_\nu\gamma^\nu P_{L}\right]\nonumber \\
 & =y_{\chi\nu}^{4}\frac{tu-m_{\phi}^{4}}{\left(t-m_{\chi}^{2}\right)^{2}}\label{eq:m-67}
\end{align}
with $q = p_1 - p_3$.

Given the above matrix elements, we can now calculate the total cross section via the relation~\cite{Zyla:2020zbs}
\begin{equation}
\sigma=\int\frac{|{\cal M}|^{2}}{64\pi s|\mathbf{p}_{1{\rm cm}}|^{2}}\di t\eqsp.\label{eq:m-42}
\end{equation}
Here, the integral covers the range (assuming  $m_1=m_2$ and $m_3=m_4$) $-(|\mathbf{p}_{1{\rm cm}}|+|\mathbf{p}_{3{\rm cm}}|)^{2}\leq t\leq-(|\mathbf{p}_{1{\rm cm}}|-|\mathbf{p}_{3{\rm cm}}|)^{2}$ with $\mathbf{p}_{i{\rm cm}}$ ($i=1$, $3$) being the spatial part of the momentum of the $i$-th particle in the center-of-mass frame, i.e.~$|\mathbf{p}_{i{\rm cm}}|=\sqrt{s/4-m_{i}^{2}}$. By applying eq.~\eqref{eq:m-42} to \eqref{eq:m-40} and \eqref{eq:m-67}, we thus obtain for the $t$-channel scenario
\begin{align}
\sigma_{\nu\bar{\nu}\rightarrow\chi\bar{\chi}}^{(t)}&=\frac{y_{\chi\nu}^{4}}{16\pi s^{2}}\left[\frac{\sqrt{s^{2}-4m_{\chi}^{2}s}\left(2\delta m^{4}+m_{\phi}^{2}s\right)}{\delta m^{4}+m_{\phi}^{2}s}-4\delta m^{2}\acoth \frac{s + 2\delta m^2}{\sqrt{s^{2}-4m_{\chi}^{2}s}} \right]\eqsp,\label{eq:m-43} \\
\sigma_{\nu\bar{\nu}\rightarrow\phi\phi^*}^{(t)} &=\frac{y_{\chi\nu}^{4}}{8\pi s^{2}}\left[-\sqrt{s^2-4m_{\phi}^{2}s}+\left(s-2\delta m^{2}\right)\acoth \frac{s-2\delta m^{2}}{\sqrt{s^2-4m_{\phi}^2s}} \right]
\end{align}
with $\delta m^{2}\equiv m_{\phi}^{2}-m_{\chi}^{2}$. 

For the $s$-channel diagram, we instead have to replace $u$ with $s$ and $t$ according to the relation $s+t+u=\sum_{i}m_{i}^{2}$, which yields
\begin{equation}
\sigma^{(s)}_{\nu\bar{\nu}\rightarrow\chi\bar{\chi}}=\frac{y_{\chi}^{2}y_{\nu}^{2}}{12\pi}\frac{s-m_{\chi}^{2}}{(s-m_{\phi}^{2})^2+(m_{\phi}\Gamma_{\phi}^{(s)})^2}\sqrt{1-\frac{4m_{\chi}^{2}}{s}}\eqsp.\label{eq:m-44}
\end{equation}

The large $s$ limits of these cross sections can be computed by noting that $\acoth(1+x)\approx\frac{1}{2}\log\left(\frac{2}{x}\right)+\frac{x}{4}-\frac{x^{2}}{16}+{\cal O}(x^{3})$
for $0<x\ll1$, which implies
\begin{align}
\lim_{s\rightarrow\infty}\sigma^{(t)}_{\nu\bar{\nu}\rightarrow\chi\bar{\chi}} & =\frac{y_{\chi\nu}^{4}}{16\pi s}\eqsp,\label{eq:m-6}\\
\lim_{s\rightarrow\infty}\sigma_{\nu\bar{\nu}\rightarrow\phi\phi^*}^{(t)} & =\frac{y_{\chi\nu}^{4}}{16\pi s}\left[\log\left(\frac{s}{m_{\chi}^{2}}\right)-2\right]\eqsp,\\
\lim_{s\rightarrow\infty}\sigma^{(s)}_{\nu\bar{\nu}\rightarrow\chi\bar{\chi}} & =\frac{y_{\chi}^{2}y_{\nu}^{2}}{12\pi s}\eqsp.
\end{align}

\subsection{Decay}
Next, we compute the decay widths of $\phi$ in both scenarios. We find that the squared amplitudes are given by
\begin{align}
|{\cal M}_{\phi\rightarrow\chi\bar{\nu} }^{(t)}|^{2} & =\sum_{{\rm spins}}y_{\chi\nu}^{4}|\bar{u}_{2}P_{L}v_{3}|^{2}\eqsp,\label{eq:m-45}\\
|{\cal M}_{\phi\rightarrow X\bar{X}}^{(s)}|^{2} & =\sum_{{\rm spins}}\frac{1}{3}\sum_{\epsilon}y_{X}^{2}|\bar{u}_{2}\gamma^{\mu}P_{L}v_{3}\epsilon_{\mu}|^{2}\label{eq:m-46}
\end{align}
with $X\in\{\nu, \chi\}$, $\bar{X} \in \{\bar{\chi}, \phi^*\}$ and an additional factor $1/3$ from the vector polarization sum. Then by evaluating the traces, we obtain
\begin{align}
|{\cal M}_{\phi\rightarrow\chi\bar{\nu}}^{(t)}|^{2} & =y_{\chi\nu}^{2}{\rm tr}[u_{2}\bar{u}_{2}P_{L}v_{3}\bar{v}_{3}P_{R}]=y_{\chi\nu}^{2}(2p_{2}\cdot p_{3}) =y_{\chi\nu}^{2}\left(m_{\phi}^{2}-m_{\chi}^{2}\right),\label{eq:m-47}\\
|{\cal M}_{\phi\rightarrow X\bar{X}}^{(s)}|^{2} & =\frac{1}{3}y_{X}^{2}\left(\frac{q_{\mu}q_{\nu}}{m_{\phi}^{2}}-g_{\mu\nu}\right){\rm tr}[u_{2}\bar{u}_{2}\gamma^{\mu}P_{L}v_{3}\bar{v}_{3}\gamma^{\nu}P_{L}]\nonumber \\
 & =\frac{1}{3}y_{X}^{2}\frac{4(p_{2}\cdot q)(p_{3}\cdot q)-2q^{2}(p_{2}\cdot p_{3})}{m_{\phi}^{2}}+\frac{4}{3}y_{X}^{2}p_{2}\cdot p_{3}\nonumber\\ & =\frac{2}{3}y_{X}^{2}\left(m_{\phi}^{2}-m_{X}^{2}\right).\label{eq:m-48}
\end{align}
Using these matrix elements, we can then obtain the decay width via the relation
\begin{equation}
\Gamma=\frac{1}{32\pi^{2}}\int|{\cal M}|^{2}\frac{|\mathbf{p}_{2{\rm cm}}|}{m_{\phi}^{2}}\di\Omega_{2}=\frac{|{\cal M}|^{2}|\mathbf{p}_{2{\rm cm}}|}{8\pi m_{\phi}^{2}}\label{eq:m-49}
\end{equation}
with $\Omega_{2}$ being the solid angle of particle 2, while $|\mathbf{p}_{2{\rm cm}}|$ is given by
\begin{equation}
|\mathbf{p}_{2{\rm cm}}|=\frac{\sqrt{m_{1}^{4}+m_{2}^{4}+m_{3}^{4}-2m_{2}^{2}m_{1}^{2}-2m_{3}^{2}m_{1}^{2}-2m_{2}^{2}m_{3}^{2}}}{2m_{1}}\eqsp.\label{eq:m-50}
\end{equation}
For $\phi\rightarrow\chi\bar{\nu}$ and $\phi\rightarrow X\bar{X}$,
we have $|\mathbf{p}_{2{\rm cm}}|=(m_{\phi}^{2}-m_{\chi}^{2})/(2m_{\phi})$
and $\sqrt{m_{\phi}^{2}/4-m_{X}^{2}}$, respectively. Finally, by ubstituting eqs.~\eqref{eq:m-47} and \eqref{eq:m-48} into~\eqref{eq:m-49},
we obtain
\begin{align}
\Gamma_{\phi\rightarrow\chi\bar{\nu}}^{(t)} & =\frac{y_{\chi\nu}^{2}\left(m_{\phi}^{2}-m_{\chi}^{2}\right){}^{2}}{16\pi m_{\phi}^{3}}\eqsp,\label{eq:m-51}\\
\Gamma_{\phi\rightarrow\nu\bar{\nu}}^{(s)} & =\frac{y_{\nu}^{2}m_{\phi}}{24\pi}\eqsp,\label{eq:m-52}\\
\Gamma_{\phi\rightarrow\chi\bar{\chi}}^{(s)} & =\frac{y_{\chi}^{2}}{24\pi}\left(1-\frac{m_{\chi}^{2}}{m_{\phi}^{2}}\right)\sqrt{m_{\phi}^{2}-4m_{\chi}^{2}}\eqsp.\label{eq:m-53}
\end{align}

\section{Sommerfeld enhancement}
\label{sec:Sommerfeld}

\begin{figure}
\centering

\includegraphics[width=0.7\textwidth]{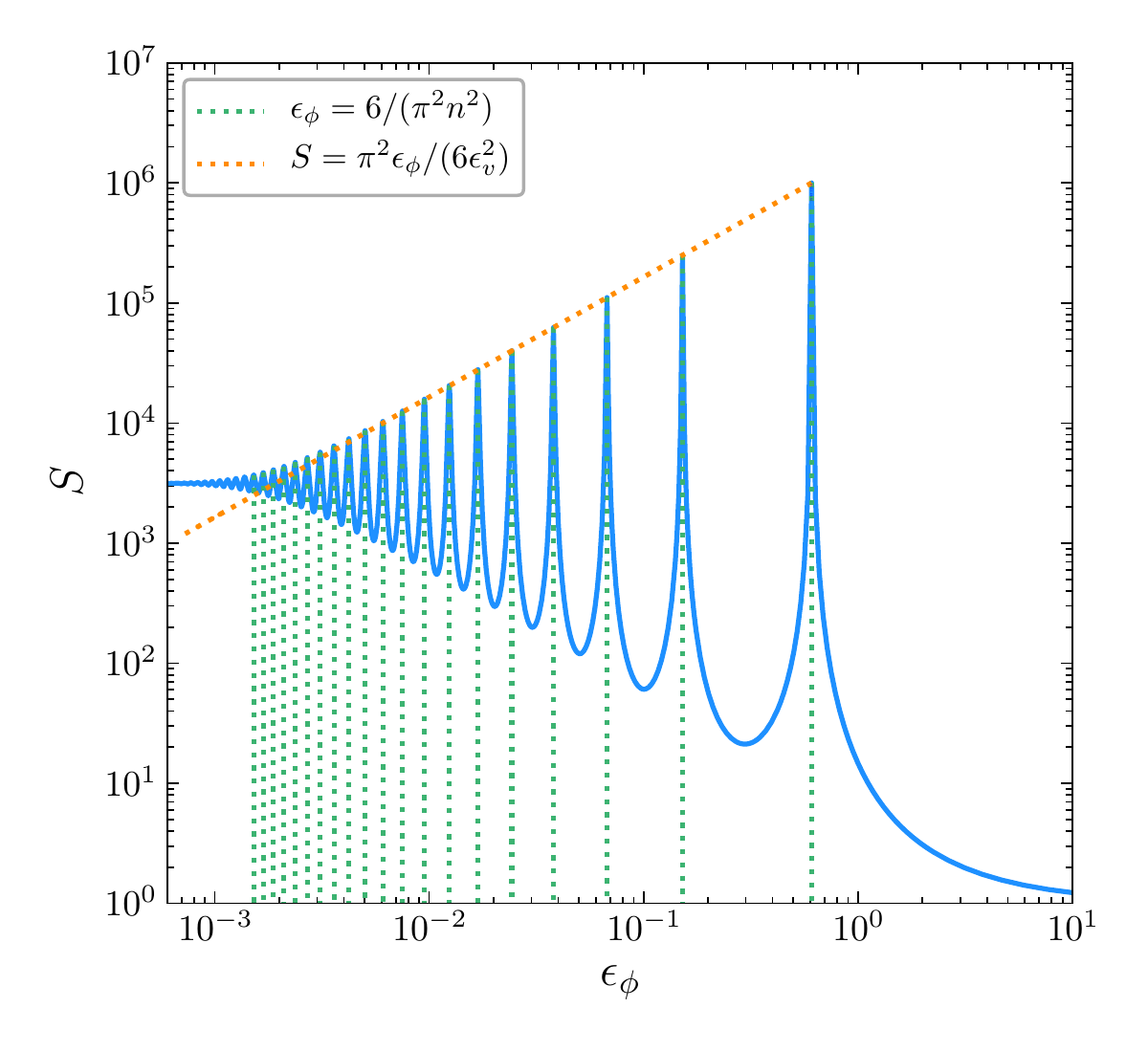}

\caption{\label{fig:Sommerfeld}The Sommerfeld enhancement factor $S$ as a function of $\epsilon_{\phi}$. The blue solid curve is obtained from eq.~\eqref{eq:m-85}
with $\epsilon_{v}=10^{-3}$. Dotted curves
show the peak positions indicated by eqs.~\eqref{eq:m-87} and \eqref{eq:m-88}.}

\end{figure}

In the presence of a light mediator, DM annihilation in the non-relativistic regime can be significantly enhanced by an effect known as  Sommerfeld enhancement~\cite{Sommerfeld}. In general, this effect can be taken into account by multiplying the cross section with a Sommerfeld enhancement
factor, which in the Hulthen potential approximation has the following analytic form~\cite{Cassel:2009wt,Slatyer:2009vg,Feng:2010zp}:
\begin{equation}
S\simeq\frac{\pi\sinh\left(\frac{12\epsilon_{v}}{\pi\epsilon_{\phi}}\right)}{\epsilon_{v}\left[\cosh\left(\frac{12\epsilon_{v}}{\pi\epsilon_{\phi}}\right)-\cos\left(\frac{12\epsilon_{v}}{\pi\epsilon_{\phi}}\sqrt{\frac{\pi^{2}\epsilon_{\phi}}{6\epsilon_{v}^{2}}-1}\right)\right]}\eqsp,\label{eq:m-85}
\end{equation}
with
\begin{equation}
\epsilon_{v}\equiv\frac{4\pi}{y_{\chi}^{2}}v\,\qquad\text{and}\qquad \epsilon_{\phi}\equiv\frac{4\pi}{y_{\chi}^{2}}\frac{m_{\phi}}{m_{\chi}}\eqsp.\label{eq:m-86}
\end{equation}
In this expression, $v$ is the relative velocity between $\chi$ and $\bar{\chi}$. A noteworthy feature of the so-defined $S$ factor is that it has local maxima at
\begin{equation}
\epsilon_{\phi}^{{\rm peak}}=\frac{6}{\pi^{2}n^{2}}\qquad\text{for}\quad n=1,\ 2,\ 3,\ \dots\label{eq:m-87}
\end{equation}
with a height that is approximately given by
\begin{equation}
S^{{\rm peak}}\simeq\frac{\pi^{2}\epsilon_{\phi}}{6\epsilon_{v}^{2}}=\left(\frac{y_{\chi}^{2}}{4\pi}\right)^{2}\frac{1}{n^{2}v^{2}}\qquad\text{for}\quad n=1,\ 2,\ 3,\ \dots\label{eq:m-88}
\end{equation}
In fig.~\ref{fig:Sommerfeld}, we plot this factor as a function of $\epsilon_{\phi}$, which illustrated that eqs.~\eqref{eq:m-87} and \eqref{eq:m-88}
are rather accurate approximation of the peak position for small $n$.

The $S$ factor in eq.~\eqref{eq:m-85} is velocity-dependent. In order
to apply this formalism to DM annihilation in the Galaxy, one can define the velocity-averaged Sommerfeld enhancement factor via~\cite{Feng:2010zp}
\begin{equation}
\bar{S}\simeq\frac{1}{v_{0}^{3}{\cal N}}\sqrt{\frac{2}{\pi}}\int_{0}^{v_{{\rm esc}}}S(v)e^{-v^{2}/(2v_{0}^{2})}v^{2}\di v\eqsp,\label{eq:m-89}
\end{equation}
with the escape velocity $v_{{\rm esc}}$, the mean velocity $v_{0}$, and
\begin{equation}
{\cal N}\equiv\text{erf}\left(\frac{v_{{\rm esc}}}{\sqrt{2}v_{0}}\right)-\sqrt{\frac{2}{\pi}}\frac{v_{{\rm esc}}}{v_{0}}e^{-v_{{\rm esc}}^{2}/(2v_{0}^{2})}\eqsp.\label{eq:m-90}
\end{equation}
Taking $v_{0}=220\,\mathrm{km/s}$ and $v_{{\rm esc}}=550\,\mathrm{km/s}$, the peaks from eqs.~\eqref{eq:m-87} and \eqref{eq:m-88} are then located at
\begin{equation}
\bar{S}^{{\rm peak}}\simeq\left(\frac{y_{\chi}^{2}}{4\pi}\right)^{2}\frac{2.04\times10^{6}}{n^{2}}\eqsp.\label{eq:m-91}
\end{equation}

\acknowledgments

We thank Laura Lopez-Honorez and Julian Heeck for useful discussions on Lyman-$\alpha$ bounds and warm dark matter, and Iason Baldes for pointing out the potentially important role of Sommerfeld enhancement in our work.
This work is supported by the ``Probing dark matter with neutrinos'' ULB-ARC convention and by the F.R.S./FNRS under the Excellence of Science (EoS) project No.\ 30820817 - be.h ``The $H$ boson gateway to physics beyond the Standard Model''.

\newpage

\bibliography{ref}
\bibliographystyle{JHEP}

\end{document}